\documentclass{aa}
\usepackage{upgreek}
\usepackage{graphicx}
\usepackage[varg]{txfonts}
\usepackage{natbib}
\usepackage{soul}
\newlength{\linwx}
\usepackage{color}

\begin{document}

\title{Pebble-isolation mass --- scaling law and implications for the formation of super-Earths and gas giants}
\author{
Bertram Bitsch \inst{1},
Alessandro Morbidelli \inst{2},
Anders Johansen \inst{1},
Elena Lega \inst{2},
Michiel Lambrechts \inst{1},
\and
Aur\'{e}lien Crida \inst{2,3}
}
\offprints{B. Bitsch,\\ \email{bert@astro.lu.se}}
\institute{
Lund Observatory, Department of Astronomy and Theoretical Physics, Lund University, 22100 Lund, Sweden
\and
University Nice-Sophia Antipolis, CNRS, Observatoire de la C\^{o}te d'Azur,Laboratoire LAGRANGE, CS 34229, 06304 Nice cedex 4, France
\and
Institut Universitaire de France, 103 Boulevard Saint-Michel, 75005 Paris, France
}
\abstract{The growth of a planetary core by pebble accretion stops at the so-called pebble isolation mass, when the core generates a pressure bump that traps drifting pebbles outside its orbit. The value of the pebble isolation mass is crucial in determining the final planet mass. If the isolation mass is very low, gas accretion is protracted and the planet remains at a few Earth masses with a mainly solid composition. For higher values of the pebble isolation mass, the planet might be able to accrete gas from the protoplanetary disc and grow into a gas giant. Previous works have determined a scaling of the pebble isolation mass with cube of the disc aspect ratio. Here, we expand on previous measurements and explore the dependency of the pebble isolation mass on all relevant parameters of the protoplanetary disc. We use 3D hydrodynamical simulations to measure the pebble isolation mass and derive a simple scaling law that captures the dependence on the local disc structure and the turbulent viscosity parameter $\alpha$. We find that small pebbles, coupled to the gas, with Stokes number $\tau_{\rm f}<0.005$ can drift through the partial gap at pebble isolation mass. However, as the planetary mass increases, particles must be decreasingly smaller to penetrate the pressure bump. Turbulent diffusion of particles, however, can lead to an increase of the pebble isolation mass by a factor of two, depending on the strength of the background viscosity and on the pebble size. We finally explore the implications of the new scaling law of the pebble isolation mass on the formation of planetary systems by numerically integrating the growth and migration pathways of planets in evolving protoplanetary discs. Compared to models neglecting the dependence of the pebble isolation mass on the $\alpha$-viscosity, our models including this effect result in higher core masses for giant planets. These higher core masses are more similar to the core masses of the giant planets in the solar system.
}
\authorrunning{Bitsch et al.}\maketitle

\section{Introduction}
\label{sec:Introduction}

Protoplanetary discs consist of gas and approximately $1\%$ dust grains. These grains collide and grow to millimeter and even
centimeter sizes \citep{2008A&A...480..859B, 2010A&A...513A..56G}.
These particles are often referred to as {\it \textup{pebbles}}. Pebbles interact with the gas disc through gas drag and drift inwards \citep{1977MNRAS.180...57W, 2008A&A...480..859B}. Pebbles can become concentrated in pressure bumps and through the streaming instability (see \citealt{Johansen2014} for a review), leading to planetesimal formation by gravitational collapse of the filaments. Planetesimals formed by the streaming instability have characteristic sizes of 100 km \citep{Johansen2015, 2016ApJ...822...55S}.

In classical planet formation models, the cores of the giant planets form through mutual collisions between these planetesimals \citep{1996Icar..124...62P}. However, to achieve a core mass high enough ($\approx$10 ${\rm M}_{\rm E}$) to attract a gaseous envelope, a surface density of planetesimals of a few times the Minimum Mass Solar Nebula (MMSN) is needed. Additionally, the growth timescale increases steeply with orbital distance, making the formation of the ice giants in the solar system basically impossible to achieve with planetesimal accretion alone. Gravitational stirring of the planetesimals by a set of growing protoplanets decreases the growth rates even more \citep{2010AJ....139.1297L}.

In recent years, a new paradigm of solid accretion has emerged: pebble accretion \citep{2010MNRAS.404..475J, 2010A&A...520A..43O, 2012A&A...544A..32L, 2012A&A...546A..18M}. When a pebble enters the planetary Hill sphere, it is subject to gas drag, which robs the pebble of angular momentum, resulting in an inward drift of the pebble onto the planet. When the largest planetesimals have grown to a few hundred kilometers in size by accreting other planetesimals, rapid pebble accretion allows further growth to cores of ten Earth masses well within the lifetime of the protoplanetary disc \citep{2016A&A...591A..72I, 2016A&A...586A..66V, Johansen2017}.

A growing planet opens a partial gap in the protoplanetary gas disc, which influences the motion of solids in the disc \citep{2006A&A...453.1129P, 2006MNRAS.373.1619R}. Pebble accretion stops when the gap carved by the planet generates a pressure maximum outside of its orbit,  which stops the inward flux of pebbles \citep{2012A&A...546A..18M, 2014A&A...572A..35L}. This is referred to as the {\it \textup{pebble-isolation mass}. As the influx of pebbles is stopped, the planet's gas envelope loses its hydrostatic support, and the envelope can then contract to form a planet with an extensive gaseous atmosphere \citep{2014A&A...572A..35L}.} In the solar system, pebble isolation is a potential mechanism to explain the dichotomy between the ice and gas giants, with ice giants never reaching the pebble-isolation mass and hence not able to undergo gas accretion within the lifetime of the protoplanetary disc \citep{2014A&A...572A..35L, 2017ApJ...848...95V, 2017AJ....154...98F}.

\citet{2014A&A...572A..35L} used hydrodynamical simulations to infer a pebble-isolation mass given by
\begin{equation}
\label{eq:Misolation}
 M_{\rm iso}^{\rm L14} \approx 20  \left( \frac{H/r}{0.05}\right)^3 {\rm M}_{\rm E} \ ,
\end{equation}
where $H/r$ is the discs aspect ratio. \citet{2014A&A...572A..35L} found a weak dependence on the viscosity parameter. This is not surprising, since a dependence of gap opening on viscosity was also reported by \citet{2006Icar..181..587C}. However, because the dependence on the viscosity was not explored in detail in \citet{2014A&A...572A..35L}, an explicit mapping of the pebble-isolation mass as a function of viscosity is of crucial importance. Additionally, the headwind felt by the particles depends on both the disc aspect ratio, as evident in Eq.~\ref{eq:Misolation} above, and on the radial, initially unperturbed, pressure gradient $\partial \ln P/\partial \ln r$.

Here, we investigate the dependence of the pebble-isolation mass on all local disc parameters, namely the disc aspect ratio $H/r$, the viscosity $\nu$, and the pressure gradient of the disc $\partial \ln P/\partial \ln r$. In order to probe this large parameter space, we adopt 3D isothermal simulations executed with the FARGOCA code \citep{2014MNRAS.440..683L, 2014A&A...564A.135B}. As particles with different sizes are coupled in different ways to the gas disc, we additionally integrate the trajectories of single pebbles with various sizes (and therefore Stokes numbers) in the gas disc to probe which particle sizes can be trapped in the pressure bump as a function of planet mass.

The paper is organised as follows. In section~\ref{sec:hydro} we present our hydrodynamical set-up and discuss the different parameters that influence the pebble-isolation mass, to which we provide a fit in absence of turbulent diffusion. In section~\ref{sec:pebbles} we integrate the trajectories of single pebbles in discs with embedded planets and infer the pebble sizes that are trapped in the pressure bumps. We then discuss turbulent diffusion of dust particles through the pressure bump. We also present the fitting formula for the pebble-isolation mass including turbulent diffusion, which is useful for planet formation simulations involving pebble accretion, in section~\ref{subsec:pebbleisolation}. In section~\ref{sec:formation} we show the influence of the new-found pebble-isolation mass on simulations of planet formation, where we compare our results to \citet{2015A&A...582A.112B}. We additionally
discuss implications of our results in section~\ref{sec:discuss} and finally summarise in section~\ref{sec:summary}.

\section{Hydrodynamic simulations}
\label{sec:hydro}

\subsection{Simulation set-up}

In order to simulate the 3D disc-planet interaction, we used the 3D hydrodynamical code FARGOCA \citep{2014MNRAS.440..683L, 2014A&A...564A.135B} in a locally isothermal configuration, where the radial temperature profile remains fixed throughout the simulation. We used the locally isothermal configuration because it allows a fast probing of parameter space in $\alpha$, $H/r$, $\partial\ln P/\partial\ln r,$ and planetary masses, which is needed to constrain the pebble-isolation mass. Here $\alpha$ is related to the viscosity through $\nu=\alpha H^2 \Omega_{\rm K}$ \citep{1973A&A....24..337S}, where $\Omega_{\rm K}$ denotes the Keplerian rotation. Additionally, locally isothermal simulations allow an easier probing of the aspect ratio because it can be set as an input parameter in contrast to simulations with heating and cooling, where the aspect ratio is set ultimately by the opacity profile that determines the cooling. Nevertheless, in section~\ref{subsec:raddiff} we test the predictions made with the isothermal simulations against simulations with heating and cooling.

We used for our simulations a 3D grid in spherical coordinates ($r, \phi, \theta$) with $315$, $720,$ and $32$ grid cells. Our grid ranged from $0.4$ to $2.5$ in radius, where the planet is located at $1$, and spanned the full azimuthal range. We used evanescent boundary conditions for the radial boundaries to damp out the spiral waves exerted by the planet in order to avoid disturbances due to reflections of the spiral waves caused by the planet. We simulated different values of the viscosity parameter $\alpha$, the aspect ratio $H/r$, the pressure gradient $\partial\ln P/\partial\ln r,$ and a range of planetary masses ($5$-$120$ ${\rm M}_{\rm E}$). 

The planetary potential was modelled with a cubic potential \citep{2009A&A...506..971K}. We used a smoothing length of $r_{\rm sm} =0.6 r_{\rm H}$, where $r_{\rm H}$ denotes the planetary Hill radius. This is the same smoothing length as in \citet{2014A&A...572A..35L}. The smoothing length has no influence on our results because the pressure bump generated by the planet outside of its orbit lies well beyond the smoothed zones. We tested different smoothing lengths, for
instance, $r_{\rm sm}=0.8 r_{\rm H}$ and $r_{\rm sm}=0.4 r_{\rm H}$, and the pressure bump outside of the planetary orbit did not change compared to our standard smoothing length of $r_{\rm sm} =0.6 r_{\rm H}$.

\subsection{Measurement of the pebble-isolation mass}

Pebbles in protoplanetary discs are subject to radial drift that
is due to the headwind they feel from the gas \citep{1977MNRAS.180...57W, 2008A&A...480..859B}. The gas orbits at a slightly sub-Keplerian speed because of the force exerted by the radial pressure gradient in the protoplanetary disc. This velocity difference is expressed as
\begin{equation}
 v_{\rm gas, \phi} = v_{\rm K} (1 - \eta) = v_{\rm K} - \eta v_{\rm K} = v_{\rm K} - \Delta v \ ,
\end{equation}
where
\begin{equation}
 \label{eq:eta}
 \eta = - \frac{1}{2} \left(\frac{H}{r}\right)^2 \frac{\partial \ln P}{\partial \ln r} \ ,
\end{equation}
and $P$ is the pressure in the protoplanetary disc. In isothermal discs the pressure is given by $P = c_{\rm s}^2 \rho_{\rm g}$, with $c_{\rm s}$ being the isothermal sound speed $c_{\rm s} = H \Omega_K$ and $\rho_{\rm g}$ the gas volume density. 

If $\eta$ is lower than $0$, the azimuthal gas velocity becomes higher than the Keplerian velocity and thus the particles feel a net outwards acceleration that will stop the inward motion of pebbles. As the planet grows, it carves a (partial) gap in the gas distribution around it by pushing material away from its orbit. This will eventually accelerate the gas outside of the planetary orbit to super-Keplerian velocities \citep{2014A&A...572A..35L}. We calculate in the following an azimuthally averaged value of $\eta$. The azimuthally averaged $\eta$ quantity gives a good handle on the generation of the pressure bump (see Appendix~\ref{ap:structure}). The planetary mass at which the created pressure bump stops the radial inward flow of pebbles through radial drift is called the pebble-isolation mass $M_{\rm iso}^\dagger$ {\it \textup{without}} diffusion.

\begin{figure}
 \centering
 \includegraphics[scale=0.7]{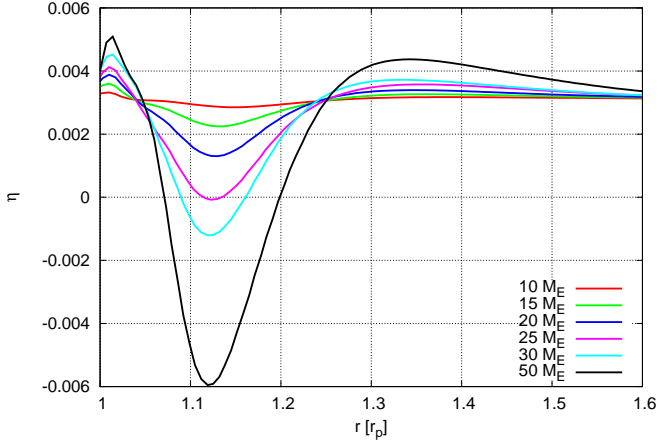}
 \caption{Pressure gradient parameter $\eta$ as a function of orbital distance from the planet for $\alpha=0.001$, $H/r=0.05,$ and different planetary masses. The location of the planet is fixed at $r=1$. A negative $\eta$ parameter indicates the formation of a pressure bump and with it super-Keplerian gas velocities that stop the inwards drift of pebbles. Here a mass of about $25$ Earth masses is needed to generate the pressure bump outside of the planetary orbit. Inside of the planetary orbit, $\eta$ is always positive.
   \label{fig:etaA001}
   }
\end{figure}

In Fig.~\ref{fig:etaA001} we display the $\eta$ parameter outside of the planetary orbit (the planet is fixed at $r=1$) for several planet masses in a disc with $H/r=0.05$, $\alpha=0.001$ and $\Sigma_{\rm g} \propto r^{-0.5}$, where $\Sigma_{\rm g}$ denotes the gas surface density of the disc. A negative value of $\eta$ means that the gas velocity is super-Keplerian, and inwards-drifting pebbles are stopped and cannot reach the planet any more. The pressure gradient is calculated for an azimuthally averaged pressure in the protoplanetary disc. In this case, a pebble-isolation mass of $M_{\rm iso}^\dagger \approx$25 ${\rm M}_{\rm E}$ is found, in rough agreement with \citet{2014A&A...572A..35L}.

\subsection{Dependence on viscosity and aspect ratio}
\label{subsec:viscHr}

\begin{figure}
 \centering
 \includegraphics[scale=0.7]{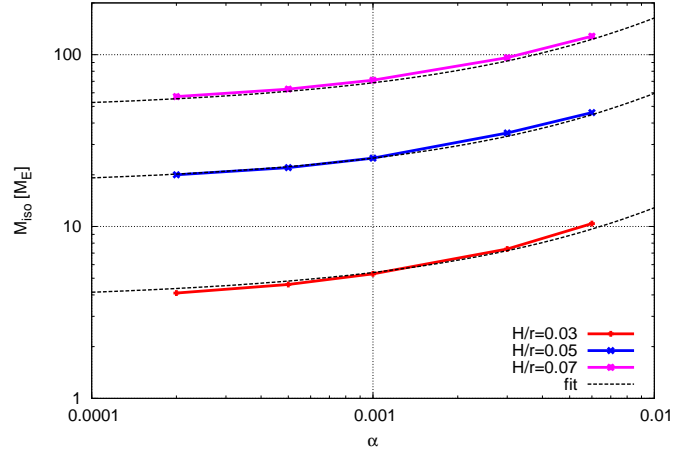}
 \caption{Pebble-isolation mass as a function of $\alpha$ and for different aspect ratios $H/r$. The pebble-isolation mass is fitted through different $\alpha$ values with a simple fit that also scales with $(H/r)^3$.
   \label{fig:Miso}
   }
\end{figure}

In Fig.~\ref{fig:Miso} we present the pebble-isolation mass determined by 3D hydrodynamical simulations with different viscosities and aspect ratios. We define the pebble-isolation mass as the planetary mass at which the pressure bump outside of the planetary orbit becomes large enough to turn $\eta$ negative. The pebble-isolation mass increases with $\alpha$ and with $H/r$, as predicted by \citet{2014A&A...572A..35L}. We show a fit for our obtained data that scales with the aspect ratio, similar to \citet{2014A&A...572A..35L},
\begin{equation}
 M_{\rm iso}^\dagger \left(H/r\right) \propto \left(\frac{H/r}{0.05}\right)^3 \ .
\end{equation}
Additionally, the fit includes a dependency on $\alpha$, which varies the pebble-isolation mass by a factor of $2-3$ between low and high $\alpha$ values. Our fit is therefore also a function of $\alpha$ in the following way:
\begin{equation}
 M_{\rm iso}^\dagger \left(\alpha \right) \propto \left( 0.34 \left(\frac{\log(\alpha_3)}{\log(\alpha)}\right)^4 + 0.66 \right)
,\end{equation}
where $\alpha_3 = 0.001$.

In contrast, the original formula from \citet{2014A&A...572A..35L}, which is at the base of the new, refined formula proposed here, is similar to the expression of the critical mass for the wake to shock in a disc \citep{2001ApJ...552..793G, 2002ApJ...572..566R}. This highlights the fact that opening a deep gap around the orbit of the planet and creating a small gap that just reverses the pressure gradient are not similar processes and obey different physics. 

To emphasise this effect, we calculated the depth of the gap following the formula for giant planet gap depths by \citet{2007MNRAS.377.1324C} given as
\begin{equation}
\label{eq:gapP}
 G(\mathcal{P}) = \left\{
  \begin{array}{cc}
   \frac{\mathcal{P}-0.541}{4} &\quad \text{if} \quad \mathcal{P}<2.4646 \\
   1.0-\exp\left(-\frac{\mathcal{P}^{3/4}}{3}\right) &\quad \text{, otherwise.}
  \end{array}
  \right. 
  \end{equation}
The parameter $\mathcal{P}$ is given by \citet{2006Icar..181..587C} as 
\begin{equation}
\label{eq:gapopen}
 \mathcal{P} = \frac{3}{4} \frac{H}{r_{\rm H}} + \frac{50}{q \mathcal{R}} \leq 1 \ .
\end{equation}
Here $q$ is the star-to-planet mass ratio, $r_{\rm H}$ the planetary Hill radius, and $\mathcal{R}$ the Reynolds number given by $\mathcal{R} = r_{\rm P}^2 \Omega_{\rm P} / \nu$. In our simulations we have checked that the left term of eq.~\ref{eq:gapopen} varies from 2 to 10 for planets that have reached the pebble-isolation mass. Thus eq.~\ref{eq:gapP} predicts a gap depth of between 15\% and 60\%, while our 3D simulations yield gap depths of only 10-20\%. The partial gap opened at the pebble-isolation mass thus corresponds to a different regime compared to the gap opening mass for giant planets of \citet{2006Icar..181..587C} and the gap depth of \citet{2007MNRAS.377.1324C}. Hence, the expression of \citet{2007MNRAS.377.1324C} for the depth of the gap should not be extrapolated to the regime of large $\mathcal{P}$ (see eq.~\ref{eq:gapopen}), low planet masses, and shallow gaps, and cannot be used to estimate the pebble-isolation mass.

\subsection{Global pressure gradient}

By changing the background gradient in surface density, the global pressure gradient changes. We varied the background surface density gradient $\Sigma_{\rm g} \propto r^{s}$ in the disc from $s=0.5$ to $s=-1.5$ and determined the pebble-isolation mass for discs with $H/r=0.05$ and $\alpha=0.001$ with the same method as above. A global inversion of the gas surface density gradient does not already imply a pressure bump in 3D simulations, in contrast to 2D simulations, because a pressure gradient inversion in 3D isothermal discs can only be reached by an inversion of the volume density gradient, where $\rho \propto r^{s-1}$ for a radially constant $H/r$. The difference between 2D and 3D simulations regarding the pebble-isolation mass is discussed in more detail in Appendix~\ref{ap:2Ddiscs}. In Fig.~\ref{fig:Misoeta} we show the pebble-isolation mass as a function of the background value of $\partial\ln P/\partial\ln r$ of the unperturbed disc.

\begin{figure}
 \centering
 \includegraphics[scale=0.7]{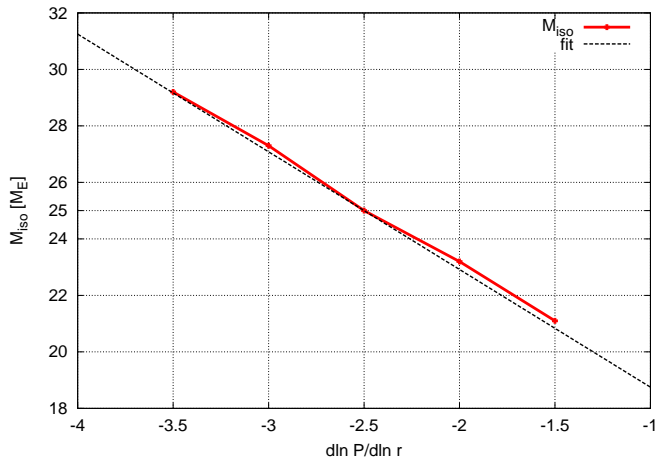}
 \caption{Pebble-isolation mass as a function of $\partial\ln P/\partial\ln r$ of unperturbed discs with different surface density gradients. Steeper surface density slopes $s$ result in more negative $\partial\ln P/\partial\ln r$ values. All simulations have been performed for planets in discs with $H/r=0.05$, $f=0$ and $\alpha=0.001$. 
   \label{fig:Misoeta}
   }
\end{figure}

The dependence on the background gradient of surface density (and thus on the background value of $\partial\ln P/\partial\ln r$) is not very strong. We approximated the dependency of the pebble-isolation mass on $\partial\ln P/\partial\ln r$ with the expression
\begin{equation}
 M_{\rm iso}^\dagger \left(\frac{\partial \ln P}{\partial \ln r}\right) \propto \left(1-\frac{\frac{\partial \ln P }{\partial \ln r} +2.5}{6} \right)
,\end{equation}
where the reference value $\partial\ln P/\partial\ln r=-2.5$ corresponds to an unperturbed disc with $H/r=$const. and $\Sigma_{\rm g} \propto r^{-0.5}$.

\subsection{Flared discs}
\label{subsec:flared}

Using $\Sigma_{\rm g} \propto r^{s}$ and $H/r \propto r^f$ ,
one can derive a dependency of $\eta$ on the orbital distance $r$ (using $P = c_{\rm s}^2 \rho = H^2 \Omega^2 \rho$),
\begin{equation}
\label{eq:etagrad}
 \eta = - \frac{1}{2} \left(\frac{H}{r}\right)^2 \frac{\partial \ln P}{\partial \ln r} = - \frac{1}{2}  \left(\frac{H_0}{r_0}\right)^2 r^{2f} (f + s -2) \ . 
\end{equation}
Here $H_0/r_0$ indicates the aspect ratio at $r=1$. This equation indicates that $\eta$ only varies radially in discs with non-constant $H/r$. As the pressure bump generated by the planet is located about $2H$ outside of the planet position, one could imagine that a change in $\eta$ with orbital distance might influence how the pressure bump is generated. We therefore tested the influence of the flaring index on the pebble-isolation mass in isothermal discs with $\alpha=0.001$, $\Sigma_{\rm g} \propto r^{-0.5}$ and $H/r=0.05r^f$, where $f$ spans from $-0.42$ to $+0.42$. We did not find any dependence on the pebble-isolation mass in discs with different flaring index. The pebble isolation mass in this case is determined {\it \textup{only}} by the local unperturbed $\partial\ln P / \partial\ln r$ value, $\alpha$ and $H/r$ at the location of the planet, but {\it \textup{not}} by the flaring of the disc itself.

\subsection{Pebble isolation mass without diffusion}

To summarise the results of Section~\ref{subsec:viscHr}-\ref{subsec:flared}, we find that the pebble-isolation mass {\it \textup{without}} diffusion $M_{\rm iso}^\dagger$ is given by
\begin{equation}
 \label{eq:MisoNEW}
  M_{\rm iso}^\dagger  = 25 f_{\rm fit} {\rm M}_{\rm E} \ ,
\end{equation}
where
\begin{equation}
\label{eq:ffit}
 f_{\rm fit} = \left[\frac{H/r}{0.05}\right]^3 \left[ 0.34 \left(\frac{\log(\alpha_3)}{\log(\alpha)}\right)^4 + 0.66 \right] \left[1-\frac{\frac{\partial\ln P}{\partial\ln r } +2.5}{6} \right] \ ,
\end{equation}
with $\alpha_3 = 0.001$.

\subsection{Radiative simulations}
\label{subsec:raddiff}

In reality, discs have complex radial temperature profiles and a non-isothermal vertical structure. Different heating sources (viscous heating, stellar heating) are balanced by radiative cooling, which can alter the disc structure quite severely compared to simple power laws \citep{2015A&A...575A..28B}. In order to test the predictions of the pebble-isolation mass (eq.~\ref{eq:MisoNEW}), we studied the pebble-isolation mass in discs with heating and cooling.

\begin{figure}
 \centering
 \includegraphics[scale=0.7]{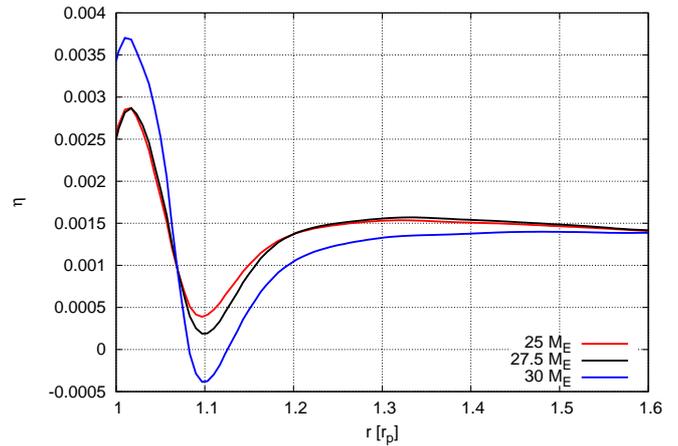}
 \caption{Pressure gradient parameter $\eta$ as a function of orbital distance in a disc set-up with heating and cooling for two different planetary masses. The planet is placed at $r=1$. The pebble-isolation mass is reached at $\approx$28.3 ${\rm M}_{\rm E}$ in the simulations, in good agreement with eq.~\ref{eq:MisoNEW}.
   \label{fig:etarad}
   }
\end{figure}

In the adiabatic (and radiative) case, the sound speed changes by a factor of $\sqrt{\gamma}$ compared to the isothermal sound speed. This leads to a difference in the scale height of the protoplanetary disc for the isothermal and adiabatic configuration, which are related in the following way:
\begin{equation}
 \label{eq:cs}
 H_{\rm adi} = \sqrt{\gamma} H_{\rm iso} \ .
\end{equation}
In the radiative configuration, we therefore used the adiabatic scale height to estimate the pebble-isolation mass in a radiative disc. The disc set-up was similar to before, where we now additionally included radiative cooling and viscous heating (with $\alpha = 6 \times 10^{-3}$) as described in \citet{2009A&A...506..971K}. At the planet location, $H_{\rm adi, pla} = 0.0414$ and $\partial\ln P/\partial\ln r=-3.26$ (the flaring index of the disc is $f=-0.38$ at this location), which leads to a pebble-isolation mass of $M_{\rm iso}^\dagger \approx$28.6 ${\rm M}_{\rm E}$ according to eq.~\ref{eq:MisoNEW}. The results of our 3D simulations in discs with heating and cooling are shown in Fig.~\ref{fig:etarad}. Clearly, eq.~\ref{eq:MisoNEW} matches the 3D simulations of discs with heating and cooling well.

\subsection{Application of the new fitting formula}

\begin{figure}
 \centering
 \includegraphics[scale=0.8]{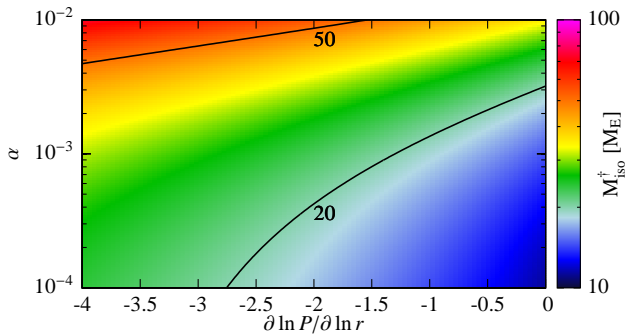} 
 \caption{Pebble-isolation mass $M_{\rm iso}^\dagger$ as a function of the pressure gradient $\partial\ln P/\partial\ln r$ and $\alpha$ in a disc with a constant $H/r=0.05$. The two black lines mark 20 and 50 ${\rm M}_{\rm E}$. Clearly, higher values of viscosity and $\partial\ln P/\partial\ln r$ result in significantly higher pebble-isolation masses.
   \label{fig:Misoalphaeta}
   }
\end{figure}

The pebble-isolation mass depends not only on the disc aspect ratio $H/r$, but also on the viscosity and the radial pressure gradient of the protoplanetary disc (eq.~\ref{eq:ffit}). For a fixed $H/r$, a change in  $\alpha$ from $10^{-4}$ to $10^{-2}$ increases the pebble-isolation mass by a factor of $\approx$3 (Fig.~\ref{fig:Miso}), while an increase in $\partial\ln P/\partial\ln r$ from $-3.5$ to $-1.5$ decreases the pebble-isolation mass by about $\approx$30\%. Making use of eq.~\ref{eq:MisoNEW}, we calculated the pebble-isolation mass in a disc with $H/r=0.05$ for different values of $\alpha$ and $\partial\ln P/\partial\ln r$ and show the resulting pebble-isolation mass in Fig.~\ref{fig:Misoalphaeta}.

Clearly, high values of viscosity ($\approx$10$^{-2}$) increase the pebble-isolation mass significantly, where the pebble-isolation mass can reach over $\approx$50 ${\rm M}_{\rm E}$ for our nominal $\partial\ln P/\partial\ln r=-2.5$. Nevertheless, the strongest
dependence of the pebble-isolation mass is on the disc aspect ratio $H/r$. The disc aspect ratio is determined by the heating of the disc, either through viscosity or stellar irradiation. As the disc evolves in time, the aspect ratio decreases in the inner part of the disc as a result of reduced viscous heating and in the outer parts as a result of a decreasing stellar luminosity \citep{2015A&A...575A..28B}. These effects reduce the pebble-isolation mass in time and only discs with high viscosities can maintain a high pebble-isolation mass in the outer parts of the disc, as the disc evolves in time.

\section{Drift of small pebbles through the bump}
\label{sec:pebbles}

Small particles ($\tau_{\rm f} \ll 1$) are strongly coupled and move with the radial gas accretion flow, while larger particles ($\tau_{\rm f} \gg 1$) are only weakly affected by gas drag. Here $\tau_{\rm f} $ denotes the Stokes number of the pebbles. The acceleration of a pebble in a gas disc is given by
\begin{equation}
 \label{eq:vpeb}
 \frac{{\rm d}\vec{v_{\rm peb}} }{{\rm d}t} = - \frac{GM_\star}{r^3} {\vec r} - 2 \Delta v \Omega_{\rm K} - \frac{1}{t_{\rm f}} \left( \vec{v_{\rm peb}} - \vec{v_{\rm gas}} \right) \ ,
\end{equation}
where ${\vec r}$ denotes the vector between the central star and the pebble, and $t_{\rm f}$ is the friction time, which is related to the Stokes number with $\tau_{\rm f} = t_{\rm f} \Omega_{\rm K}$ and $\Delta v = \eta v_{\rm K}$. The variables $\vec{v_{\rm peb}}$ and $\vec{v_{\rm gas}}$ are the pebble and gas velocities, respectively.

\begin{figure}
 \centering
 \includegraphics[scale=0.7]{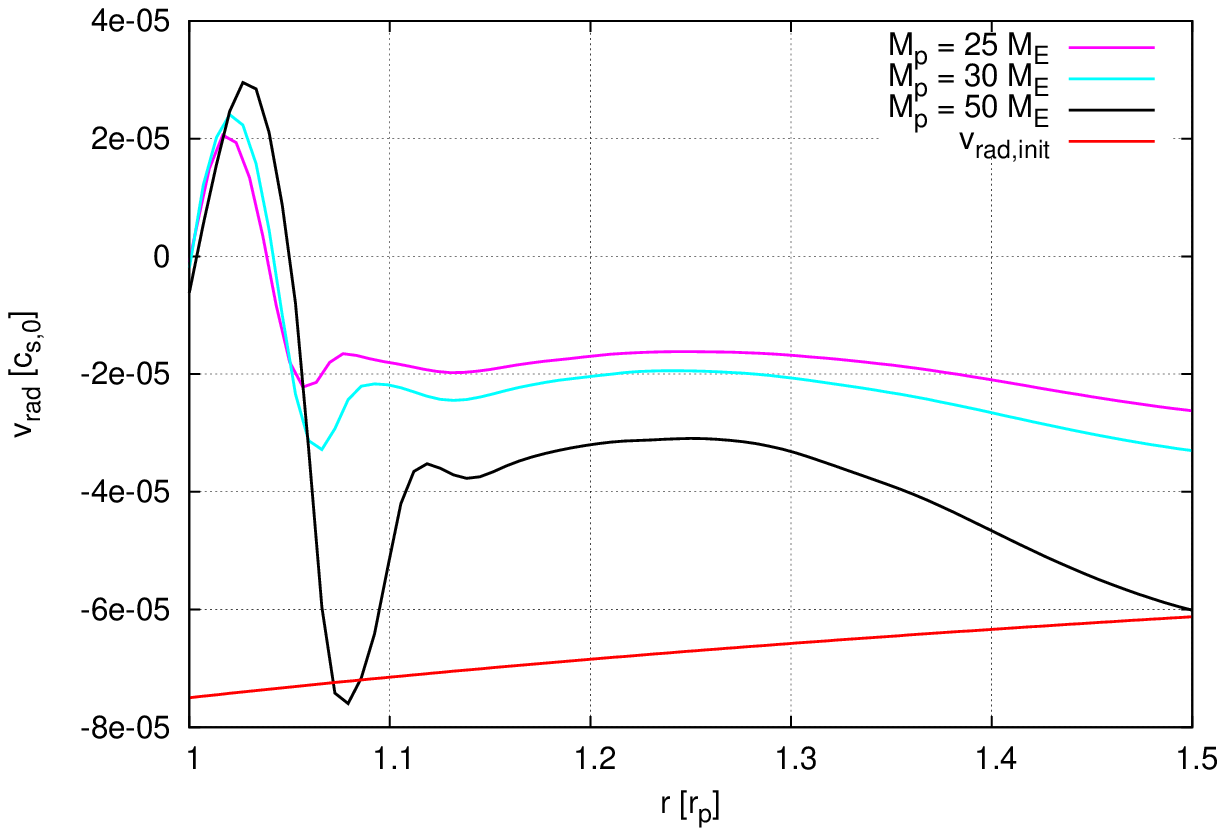} 
 \includegraphics[scale=0.7]{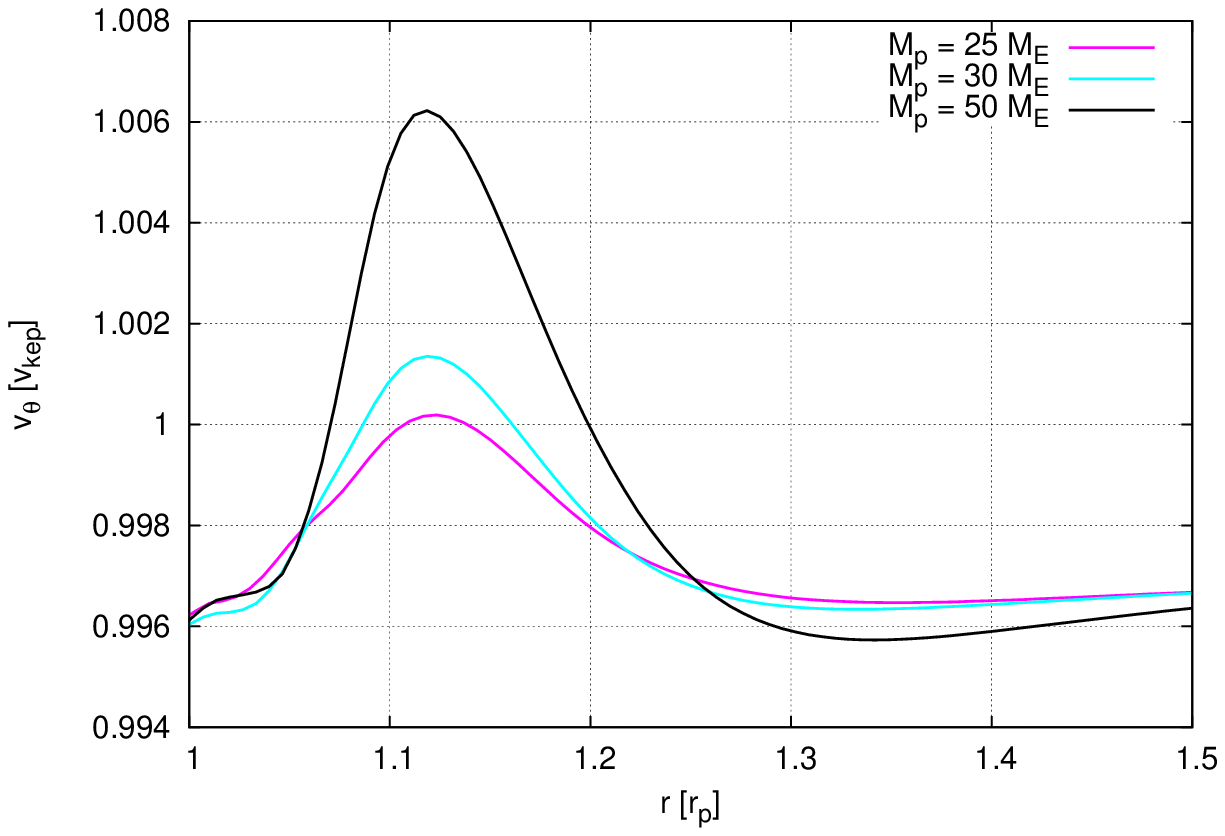}  
 \caption{Midplane gas velocities as a function of orbital distance in discs with $H/r=0.05$, $\alpha=0.001,$ and $\Sigma \propto r^{-0.5}$. The top plot shows the radial velocity in units of the sound speed at $r=1.0$. A negative velocity indicates an inward flow of the gas. The red line indicates the radial velocity of the unperturbed disc. The bottom plot shows the azimuthal velocity in units of the Keplerian velocity. If $v_\theta$ is larger than $1$, the gas orbits super-Keplerian, indicating the pressure bump in the disc.
   \label{fig:gasvel}
   }
\end{figure}

In Fig.~\ref{fig:gasvel} we show the radial and azimuthally averaged gas velocities in locally isothermal discs with embedded planets. The pressure bump generated by the planet outside of its orbit is clearly visible in the azimuthal velocity pattern, where the gas can reach speeds higher then the Keplerian value. Particles entering this pressure bump can be trapped, depending on their size.

A negative radial velocity indicates an inward flow of the gas, while a positive radial velocity indicates an outward movement of the gas. The planet generates a radial outward flow of gas close to its vicinity, but limited to the region in front of the pressure bump. This outward flow is related to the gap-opening process, where the planet pushes the material away from its orbit. This material can then move upwards to maintain hydrostatic equilibrium and again falls in from the top regions of the disc onto the planet. The same meridional flow was also observed for gap-opening giant planets \citep{2014Icar..232..266M}. In this region, inflowing particles might in principle be trapped as well.

After the hydrodynamical simulations shown in section~\ref{sec:hydro} reached an equilibrium state, we integrated the movement of test particles in the steady-state gas distribution to determine the dependence of the pebble-isolation mass on the Stokes number of the particles. The steady-state surface density profile and $\eta$ profile of the disc are shown in Appendix~\ref{ap:structure}, where the pressure bump is clearly visible for a 25 ${\rm M}_{\rm E}$ planet.

We integrated the pebble trajectories in 2D planes of the protoplanetary disc, where we mainly focused on integration in the disc midplane. Pebbles injected at higher altitudes in the disc are additionally subject to vertical settling, which moves them quickly towards the midplane. Integrating the pebble trajectories in a 2D plane above the disc midplane revealed, however, that pebbles are stopped at all altitudes when the planet has reached isolation mass. This implies that at higher altitudes pebbles cannot drift through the generated pressure bump either.

\begin{figure}
 \centering
 \includegraphics[scale=1.0]{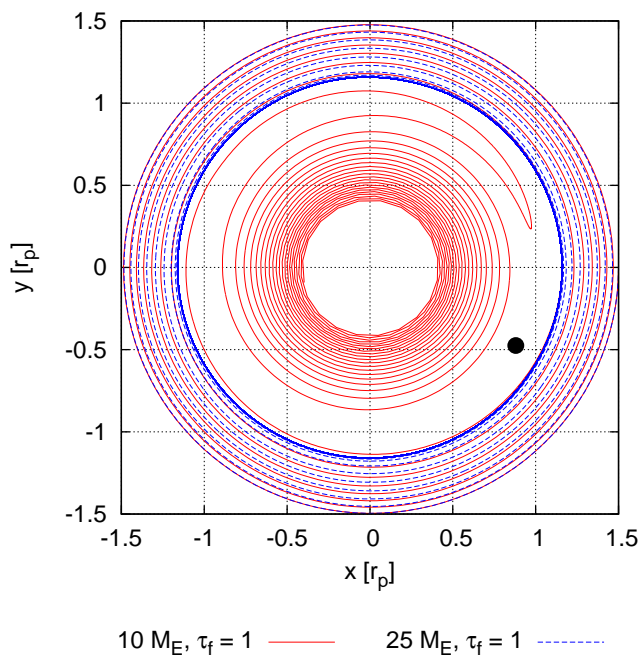} 
 \caption{Trajectories of pebbles with $\tau_{\rm f}=1.0$ that started at $(1.5;0)$ moving in the gas velocity field generated by planets with $M_{\rm P} =$ 10 ${\rm M}_{\rm E}$ and $M_{\rm P} =$ 25 ${\rm M}_{\rm E}$. We plot the pebble trajectories in the rotating frame, and the planet position is marked by the black dot. The 10 ${\rm M}_{\rm E}$ planet does not generate a pressure bump, so the pebble drifts through all the way towards the inner boundary of the computational domain. The 25 ${\rm M}_{\rm E}$ planet, on the other hand, generates a pressure bump outside of its orbit and thus blocks the flow of pebbles.
   \label{fig:pebbleSt1}
   }
\end{figure}

In Fig.~\ref{fig:pebbleSt1} we show the trajectories of pebbles in the midplane with constant $\tau_{\rm f} = 1.0$ in the gas velocity field generated by planets with 10 ${\rm M}_{\rm E}$ and 25 ${\rm M}_{\rm E}$. As the 10 ${\rm M}_{\rm E}$ planet does not generate a pressure bump outside of its orbit (Fig.~\ref{fig:etaA001}), the pebbles drift through towards the inner disc. The 25 ${\rm M}_{\rm E}$ planet generates a pressure bump in the disc (Fig.~\ref{fig:etaA001}) where the pebbles can be trapped.

\begin{figure}
 \centering
 \includegraphics[scale=0.7]{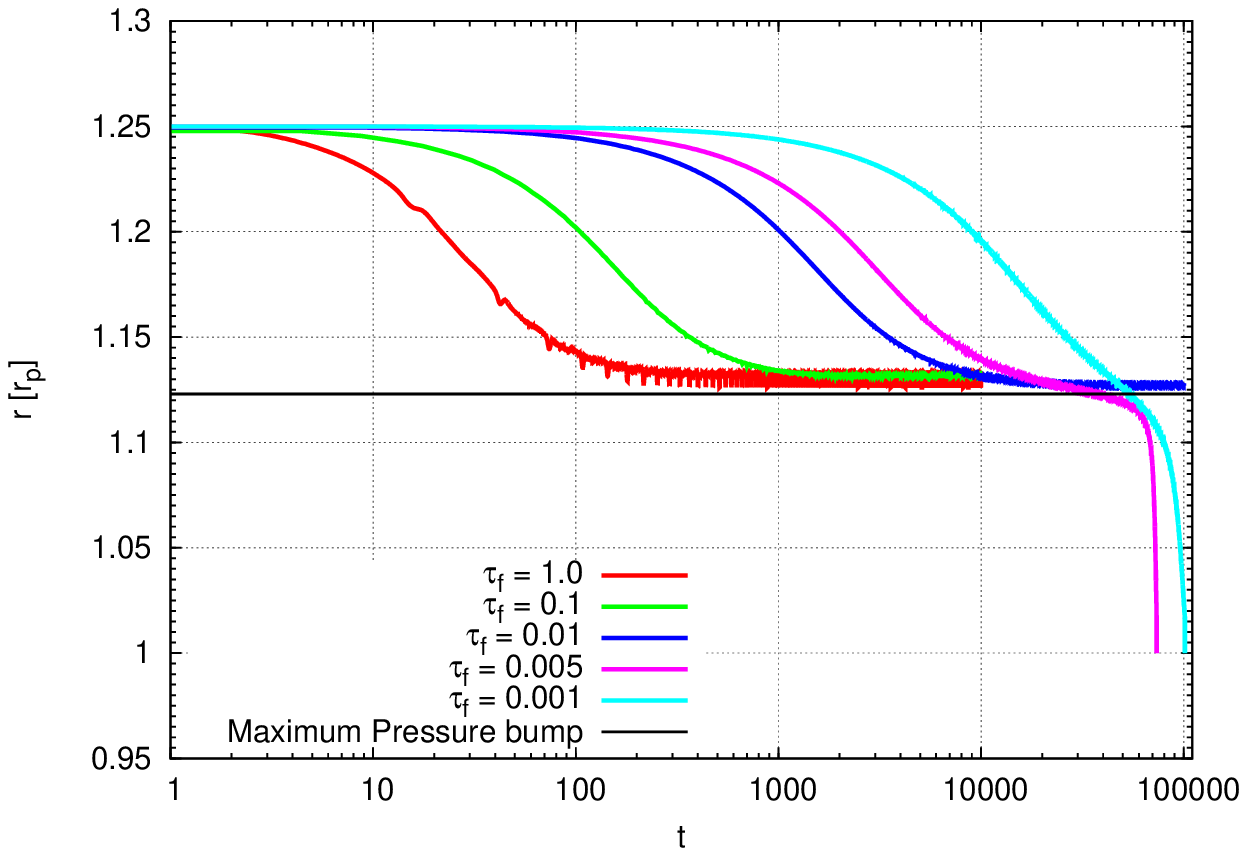} 
 \includegraphics[scale=0.7]{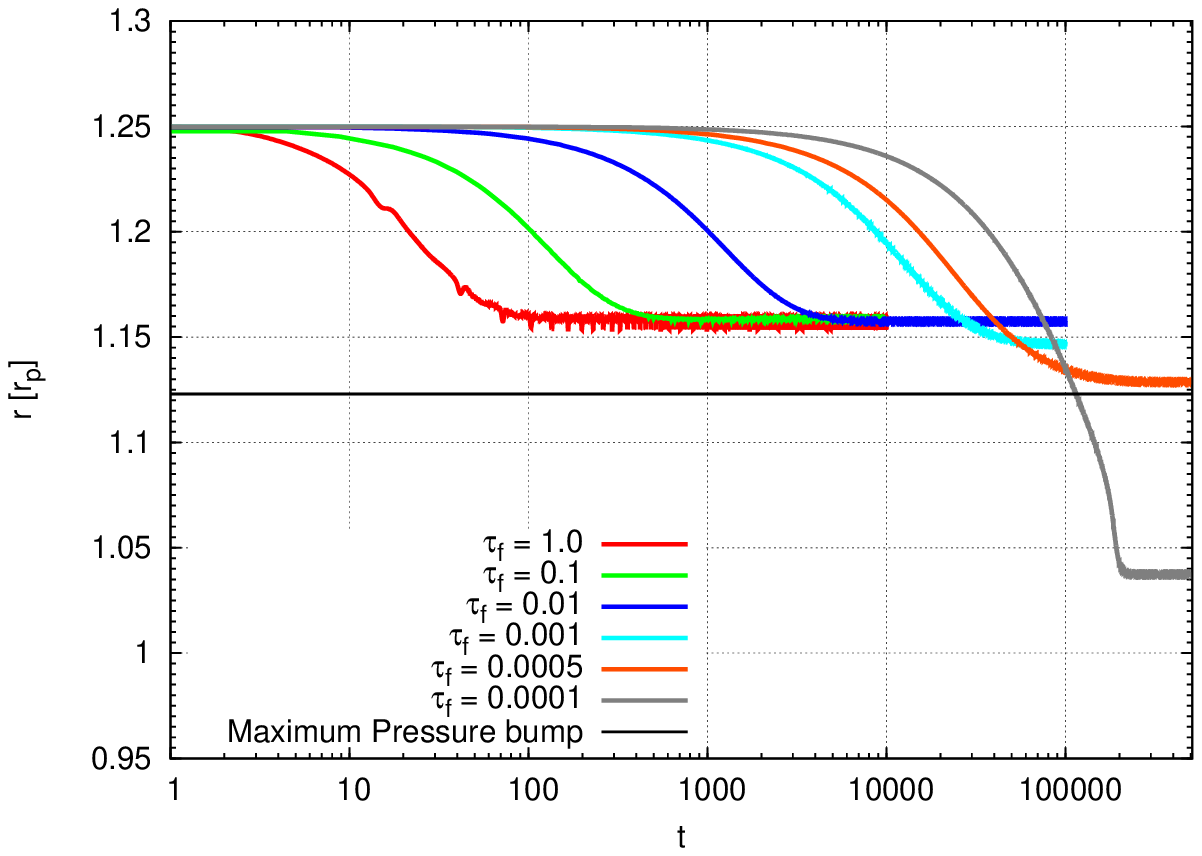}  
 \caption{Evolution of the orbital distance of pebbles as they drift through a disc with an embedded 25 ${\rm M}_{\rm E}$ planet (top) and an embedded 30 ${\rm M}_{\rm E}$ planet (bottom). Even though the planet has generated a pressure bump outside of its orbit (maximum at $r=1.123$, see Fig.~\ref{fig:etaA001}), small pebbles with $\tau_{\rm f} < 0.005$ can drift through the pressure bump for the 25 ${\rm M}_{\rm E}$ planet, while the Stokes number of the particles has to be an order of magnitude smaller to drift through the pressure bump generated by the 30 ${\rm M}_{\rm E}$ planet.
   \label{fig:DriftM25}
   }
\end{figure}

In the top panel of Fig.~\ref{fig:DriftM25}, we show the time evolution of the orbital distance of integrated pebble trajectories with different Stokes number $\tau_{\rm f}$ in a disc with an embedded 25 ${\rm M}_{\rm E}$ planet. Pebbles with $\tau_{\rm f} > 0.005$ are trapped in the pressure bump and do not drift inwards any more. Particles with $\tau_{\rm f} < 0.005$ are well enough coupled with the gas to move through the pressure bump towards the system interior to the 25 ${\rm M}_{\rm E}$ planet. The pressure bump generated by the 25 ${\rm M}_{\rm E}$ planet is even quite weak (Fig.~\ref{fig:etaA001}), which explains why small particles can still drift through.

The pressure bump generated by the planet increases in strength with planetary mass (Fig.~\ref{fig:etaA001}), which allows the trapping of pebbles with smaller $\tau_{\rm f}$ in it for higher planetary masses (bottom panel in Fig.~\ref{fig:DriftM25} for 30 ${\rm M}_{\rm E}$ planet). Increasing the planetary mass by $20\%$ reduces the Stokes number of particles that can still drift through the generated pressure bump by more than an order of magnitude. Now particles with $\tau_{\rm f}>5 \times 10^{-4}$ are trapped inside the pressure bump and can no longer reach the inner system.

Very small particles, however, with $\tau_{\rm f} \approx$1$\times 10^{-4}$, are so strongly coupled with the gas that they completely
follow the gas flow \citep{2008A&A...480..859B} and are thus no longer blocked by the pressure bump located at $r=1.13$  (see section~\ref{subsec:gasvel}). Instead, the pebbles drift through the pressure bump, but are then caught just outside of the planetary orbit because of the radial outward flow of the gas (Fig.~\ref{fig:gasvel}), which prevents further inward drift. For the 25 ${\rm M}_{\rm E}$ planet, the pebbles also drift through the radial outward flow of the gas because it was not strong enough to keep the pebbles from the inner disc (see section~\ref{subsec:gasvel}).

\subsection{Dependence on the radial gas velocity}
\label{subsec:gasvel}

The radial velocity of the gas in an $\alpha$ accretion disc is determined directly by the viscosity of the disc,
 \begin{equation}
  v_{\rm r} = - \frac{3}{2} \frac{\nu}{r} = - \frac{3}{2} \frac{\alpha H^2 \Omega_{\rm k}}{r} \ .
 \end{equation}
However, there is a strong debate in the literature about the causes of the turbulence and about the size of its magnitude \citep{Turner2014}. As the drift velocity of the particles depends on the gas velocity, we investigate in this section how a change in the radial gas velocity influences particle drift through the disc. We are particularly interested in how a change in the radial gas velocity allows or hinders particles from drifting through the pressure bump generated by the azimuthal gas velocity changes induced by the planet outside of its orbit (Fig.~\ref{fig:gasvel}). We artificially modified the radial velocity pattern to a fixed value, but kept the azimuthal gas velocity profile of a disc perturbed by a planet (Fig.~\ref{fig:gasvel}). In this way, we mimicked the effects of different levels of turbulence without simulating discs with magneto-rotational instability (MRI) or vertical-shear instability turbulence.

\begin{figure}
 \centering
 \includegraphics[scale=0.7]{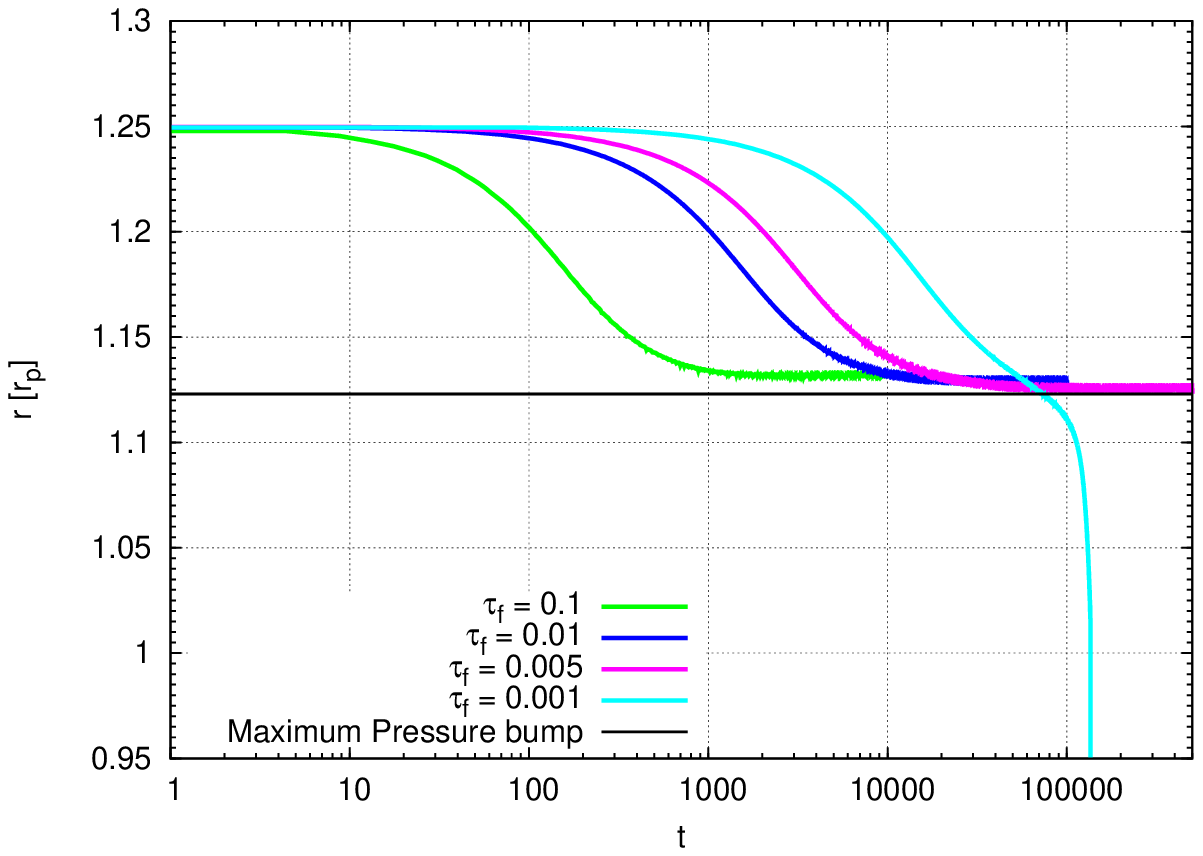} 
 \includegraphics[scale=0.7]{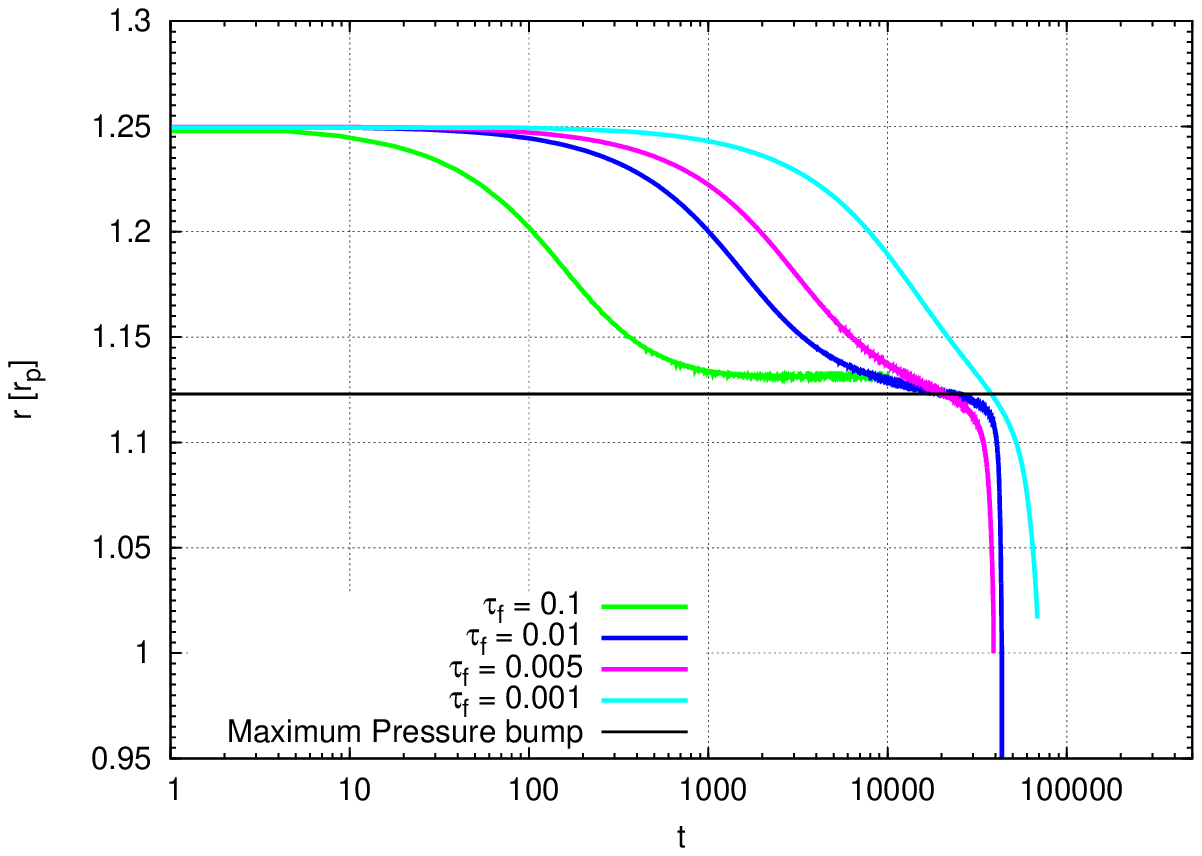}  
 \caption{Evolution of the orbital distance of pebbles as they drift through a disc with an embedded 25 ${\rm M}_{\rm E}$ planet, where the radial velocity is fixed to $-1 \times 10^{-5} c_{\rm s,0}$ (top) and $-3 \times 10^{-5} c_{\rm s,0}$. As the drift speed of the particles depends on the gas velocities, particles of different sizes cannot be blocked or drift through compared to the nominal case shown in Fig.~\ref{fig:DriftM25}. In particular, a slower gas flow allows for a more efficient particle trapping in the pressure bump generated by the planet.
   \label{fig:Driftvel}
   }
\end{figure}

In Fig.~\ref{fig:Driftvel} we show the trajectories of pebbles embedded in discs with fixed radial gas velocities, but with azimuthal gas velocity profiles that correspond to Fig.~\ref{fig:gasvel} for the 25 ${\rm M}_{\rm E}$ planet. In the top panel we show the trajectories of pebbles in discs with radial gas velocities lower than in Fig.~\ref{fig:DriftM25}, while in the bottom panel
we show the trajectories of pebbles in discs with radial gas velocities higher than in Fig.~\ref{fig:DriftM25}. Clearly, a lower radial gas velocity allows for more efficient trapping of smaller pebbles compared to a higher radial gas velocity.

The reason is that the radial pressure gradient is negative everywhere in the disc except at the centre of the outer pressure bump generated by the planet, where it is zero for a planet that has just reached pebble-isolation mass. Hence any tiny radial gas velocity can transport dust particles of any size across the pressure bump. Pebbles are only safe when the planet is more massive and the inner edge of the pressure bump has a positive radial gradient in pressure. As the radial gas velocity is determined by the viscosity in an $\alpha$-accretion disc, the movement of particles is determined by viscosity for discs with high viscosity and by drift in discs with low viscosity \citep{2016MNRAS.459L..85D}.

\begin{figure}
 \centering
 \includegraphics[scale=0.7]{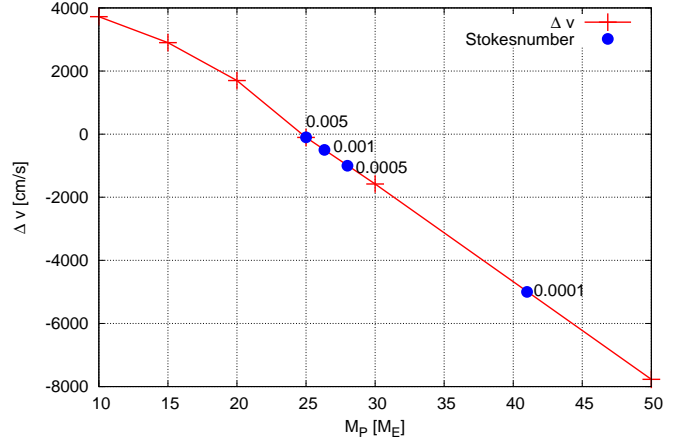} 
 \caption{Velocity perturbation $\Delta v$ as a function of planet mass at the location of the pressure bump. A negative $\Delta v$ value indicates that the gas speed is super-Keplerian. The blue circles mark the lowest Stokes number of particles that can be stopped at the pressure bump assuming a radial gas velocity of $1 \rm cm/\rm s$ and following eq.~\ref{eq:driftvelocity}.
   \label{fig:DeltavMP}
   }
\end{figure}

Pebbles with a (positive) terminal velocity
\begin{equation}
 v_{\rm r,t} = \tau_{\rm f} \frac{1}{\rho} \frac{\partial \ln P}{\partial \ln r} = 2 \tau_{\rm f} \Delta v
\end{equation}
high enough to compensate for the (negative) radial gas velocity $v_{\rm r,g}$ can be trapped in the pressure bump. This can be reformulated as 
\begin{equation}
\label{eq:driftvelocity}
 \tau_{\rm f} >  \frac{v_{\rm r,g}}{2 \Delta v} \ .
\end{equation}
Using this equation, we can estimate the Stokes number of particles that are blocked by the pressure bump generated by the planet by just looking at the velocity fields. The radial gas velocity is $\approx$1$\frac{\rm cm}{\rm s}$ in our disc model, and the measured $\Delta v$ is shown in Fig.~\ref{fig:DeltavMP}. This indicates that pebbles with $\tau_{\rm f} > 0.005$ should be blocked by a planet of 25 ${\rm M}_{\rm E}$ and particles with $\tau_{\rm f} > 0.0003$ should be blocked by a planet of 30 ${\rm M}_{\rm E}$, in agreement with our simulations (Fig.~\ref{fig:DriftM25}). We have marked the minimal Stokes number of particles that can be blocked by the pressure bump generated by the planet with blue circles in Fig.~\ref{fig:DeltavMP} for a radial velocity of $1 \frac{\rm cm}{\rm s}$. The lowest Stokes number depends linearly on the radial gas velocity, which is slightly different for the higher planetary masses (Fig.~\ref{fig:gasvel}). For a 50 ${\rm M}_{\rm E}$ planet, the radial gas velocity is roughly $\approx$2$\frac{\rm cm}{\rm s}$ (Fig.~\ref{fig:gasvel}), higher than for the 25 ${\rm M}_{\rm E}$ planet, because the planet influences the velocity pattern of the disc. This implies that pebbles with $\tau_{\rm f} > 1.2 \times 10^{-4}$ can be stopped by a 50 ${\rm M}_{\rm E}$ planet, but we note that the blue dots in Fig.~\ref{fig:DeltavMP} correspond to $v_{\rm g} = 1$ cm/s. This is also in agreement with our simulations. These results do not take diffusion into account, which we discuss in the next section.

\subsection{Diffusion of dust particles}
\label{subsec:diffusion}

Our hydrodynamical simulations do not include any turbulent motion of the gas, such as those seen in magnetohydrodynamics
(MHD) simulations \citep{2017arXiv170700729B} or in simulations with the vertical shear instability \citep{2013MNRAS.435.2610N,2016arXiv160702322S}. These turbulent motions in the gas velocities can act on the movements of the pebbles, giving them random kicks. Several authors have considered the effects of diffusion on particles embedded in discs with planets. \citet{2006A&A...453.1129P} included diffusion into the motion of dust particles in gas discs in the presence of 30 ${\rm M}_{\rm E}$ planets and estimated this effect to be of the order of $1\%$, indicating that diffusion of dust particles does not play a role in opening a gap in dust distribution of protoplanetary discs. \citet{2016A&A...585A..35P}, on the other hand, studied dust filtration by giant planets in the context of transition discs. Giant planets open deep gaps in protoplanetary discs that prevent dust from drifting through. However, in their 2D simulations, the authors found that a planet of 1 $M_{\rm Jup}$ still does not stop all dust particles and small dust grains ($\tau_f \approx$10$^{-3}$) can drift through the gap generated by the planet at $20$ AU. Their disc set-up would lead to a pebble-isolation mass of 27.5 ${\rm M}_{\rm E}$ according to eq.~\ref{eq:MisoNEW}, where the pebble-isolation mass is reached in 2D disc simulations at lower masses (see Appendix~\ref{ap:2Ddiscs})
than in 3D simulations. 

\citet{2016A&A...585A..35P} considered turbulent mixing of dust particles, where the dust diffusivity follows the prescriptions by \citet{2007Icar..192..588Y}, which depend on the Stokes number and the gas diffusivity (assumed to be equal to the disc viscosity). If then the pressure gradient is not steep enough, the particles can be released from the pressure bump, where particles with $\tau_{\rm f} > \alpha$ are trapped. Smaller particles are diffused out of the pressure bump and dragged by the gas. Without diffusion, the trapping of particles is more efficient \citep{2012A&A...545A..81P}. This mechanism allowed the small particles to move across the pressure bump generated by the planet in \citet{2016A&A...585A..35P}, while our simulations show an effective trapping of small particles as a result of the lack of diffusion. The difference of \citet{2016A&A...585A..35P} to \citet{2006A&A...453.1129P} is probably related to different prescriptions of diffusion. We therefore estimate the effects of diffusion in the following for the pebble-isolation mass.

\begin{figure}
 \centering
 \includegraphics[scale=0.8]{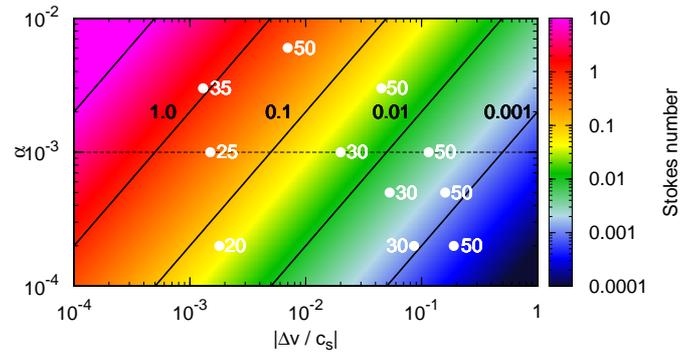} 
 \caption{Critical Stokes number of particles that are affected by turbulent diffusion as a function of absolute $\Delta v / c_{\rm s}$ in the pressure bump and $\alpha$. Particles with Stokes numbers lower than the critical value easily diffuse and can drift through a pressure bump. The black lines denote particles with Stokes numbers of 10, 1, 0.1, 0.01, and 0.001 from left to right. The white dots correspond to different planetary masses, where we have taken the maximum value of the pressure bump generated by the planet for simulations with different $\alpha$. The planetary masses are marked with the white numbers next to the dots. The horizontal dashed black line markes our nominal $\alpha=0.001$.
   \label{fig:Stokesmap}
   }
\end{figure}

The equilibrium between radial advection of dust particles and turbulent diffusion is achieved when (as also stated in \citealt{2012A&A...545A..81P})
\begin{equation}
\label{eq:diffusion}
 v_{\rm r,p} \rho_{\rm p} - D \rho_{\rm g} \frac{{\rm d} \epsilon}{{\rm d} r} = 0\ ,
\end{equation}
where $v_{\rm r,p}$ is the radial velocity of the pebbles, $\rho_{\rm p}$ the pebble density, $\rho_{\rm g}$ the gas density, $D$ the diffusion coefficient parameterised by $D = \alpha c_{\rm s} H_{\rm g}$ , and $\epsilon = \rho_{\rm p} / \rho_{\rm g}$ is
the dust-to-gas ratio. The radial velocity of the pebbles is given by
\begin{equation}
 v_{\rm r,p} = - 2 \tau_{\rm f} \Delta v \ .
\end{equation}
To efficiently trap dust in a pressure bump despite turbulent diffusion, the equilibrium flux stated above must be obtained for a radial dust scale-length ($H_{\rm p}$) that is roughly equal to the extent of the pressure bump ($H_{\rm b}$). As the latter has an approximate width of one gas scale-height, this leads to
\begin{equation}
 H_{\rm p} \sim H_{\rm g} \ .
\end{equation}
We estimate the scale-height of the dust in the pressure bump from eq.~\ref{eq:diffusion} to obtain
\begin{equation}
 \frac{D}{2 \tau_{\rm f} \Delta v} \sim H_{\rm g} \ .
\end{equation}
Expanding now $D,$ we arrive at
\begin{equation}
 \frac{\alpha c_{\rm s} H_{\rm g}}{2 \tau_{\rm f} \Delta v} \sim H_{\rm g} \ ,
\end{equation}
which leads to
\begin{equation}
\label{eq:stokes}
 \tau_{\rm f} \sim \frac{\alpha}{2 \Pi} 
,\end{equation}
where $\Pi = \Delta v / c_{\rm s}$. Particles with Stokes numbers larger than this (eq.~\ref{eq:stokes}) are still trapped in a pressure bump, particles smaller than this can diffuse through the pressure bump. We show this critical Stokes number in Fig.~\ref{fig:Stokesmap} for a disc with $H/r=0.05$.

\begin{figure}
 \centering
 \includegraphics[scale=0.7]{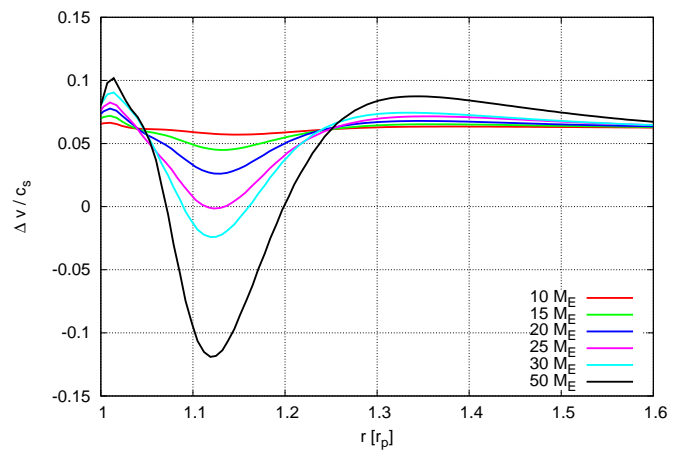} 
 \caption{$\Delta v / c_{\rm s}$ as function of orbital distance in discs with $\alpha=0.001$ and $H/r=0.05$ and embedded planets of different masses. Note that $\Delta v / c_{\rm s} = \eta / (H/r)$. The $\Delta v / c_{\rm s}$ values shown here correspond to the white dots at $\alpha=0.001$ (dashed line) in Fig.~\ref{fig:Stokesmap}.
   \label{fig:Deltav}
   }
\end{figure}

We show $\Delta v / c_{\rm s}$ as a function of planetary mass in a disc with $\alpha=0.001$ and $H/r=0.05$ in Fig.~\ref{fig:Deltav}. We can now compare directly to Fig.~\ref{fig:Stokesmap} to determine when and how pebbles inside the pressure bump are affected by diffusion. For a 30 ${\rm M}_{\rm E}$ planet in a disc with $\alpha=0.001$, the minimum Stokes number of particles that are not affected by diffusion is $\approx$0.025, while for the 50 ${\rm M}_{\rm E}$ planet, this particle size is $\approx$0.004, which corresponds to the particle size stopped by the 25 ${\rm M}_{\rm E}$ planet in the case of no diffusion (Fig.~\ref{fig:DriftM25}). This indicates that the pebble-isolation mass for these Stokes numbers in case of diffusion is a factor of $\approx$2 higher than in the case without diffusion for $H/r=0.05$ and $\alpha=0.001$, depending on $\tau_{\rm f}$.

Inside the pressure bump generated by a Jupiter-mass planet in 2D simulations of a disc with $\alpha=0.001$ and $H/r=0.05$, the maximum $\Delta v / c_{\rm s} = 0.66$. This leads to a critical Stokes number of $\tau_{\rm f} = 7.5 \times 10^{-4}$, in agreement with previous studies including diffusion, see for example \citet{2016A&A...585A..35P}, where very small particles can drift through the pressure bump generated by large planets. Adding the effects of particle diffusion in 2D disc simulations, Ataiee et al. (2017, in prep) found that diffusion can increase the pebble-isolation mass, in agreement with \citet{2012A&A...545A..81P} and our estimates.

The difference between the pebble-isolation mass derived from pure hydrodynamical simulations compared to simulations taking diffusion into account also depends on the level of turbulence in protoplanetary discs. For example, blocking particles with $\tau_{\rm f}=0.001$ requires for $\alpha=2 \times 10^{-4}$ an increase in pebble-isolation mass by about a factor of 1.5, while for $\alpha=0.001$ an increase of much more than a factor of 2 is needed (a Jupiter-mass planets blocks pebbles with $\tau_{\rm f} > 7.5 \times 10^{-4}$ in the case of diffusion).

However, the level of turbulence in discs is not very well constrained, where $\alpha$ values from $10^{-4}$ to $10^{-1}$ can be reached in simulations of the MRI (see \citealt{Turner2014} for a review). Simulations with hydrodynamical instabilities in regions of the disc that are not subject to MRI-driven turbulence show $\alpha$ values of a few times $10^{-4}$, which probably sets a lower limit on turbulence \citep{2003ApJ...582..869K, 2013MNRAS.435.2610N, 2016arXiv160702322S}. Recent simulations indicate that disc winds could be the main driver of disc accretion. In these simulations, the midplane regions only have a very low level of turbulence. For such a low level of turbulence, the effect of turbulent diffusion on the pebble-isolation mass is quite low compared to discs with high viscosity.

The dominant particle size in the planet formation simulations of \citet{2015A&A...582A.112B} with $\alpha=0.0054$ presented below is of the order of $0.1$ (Fig.~\ref{fig:r0t0Stokes}), and about $0.01-0.2$ when taking fragmentation into account (see below). From Fig.~\ref{fig:Stokesmap} it can be inferred that taking turbulent diffusion into account, the pebble-isolation mass for these Stokes numbers is higher than predicted in the case without diffusion by about a factor of 2.

\subsection{Pebble-isolation mass including diffusion}
\label{subsec:pebbleisolation}

The pebble-isolation mass does not only depend on $H/r$, $\alpha$ and $\partial\ln P/\partial\ln r$, but also, as shown in the previous subsection, on the turbulent diffusion of particles. The critical pressure gradient parameter $\Pi_{\rm crit}$ to block particles of Stokes number $\tau_{\rm f}$ is given by
\begin{equation}
 \Pi_{\rm crit} = \frac{\alpha}{2\tau_{\rm f}} \ .
\end{equation}
From our hydrodynamical simulations we can measure how $\Pi=\Delta v /c_{\rm s}$ in the pressure bump generated by a planet that has already reached $M_{\rm iso}^\dagger$ changes with planetary mass as
\begin{equation}
 \Pi = \lambda (M_{\rm p}/{\rm M}_{\rm E} - M_{\rm iso}^\dagger/{\rm M}_{\rm E}) \ ,
\end{equation}
where
\begin{equation}
 \lambda \approx 0.00476 / f_{\rm fit} \ ,
\end{equation}
where $f_{\rm fit}$ is defined in eq.~\ref{eq:ffit}. This fit only applies to planets that have already reached the pebble isolation mass without diffusion $M_{\rm iso}^\dagger$, because $\lambda$ gives the slope of the change of $\Pi$ inside the pressure bump generated by the planet, where $M_{\rm iso}^\dagger$ is the minimum mass needed to invert the radial pressure gradient $\partial \ln P /\partial \ln r$ in the disc. We note that $\lambda$ is only valid until $M_{\rm p} \approx$2.5$M_{\rm iso}^\dagger$, when $\lambda$ changes, because the growing planet slowly transitions into the  gap depth regime predicted by \citet{2007MNRAS.377.1324C}. When setting $\Pi_{\rm crit} = \Pi,$ we can define the pebble isolation mass {\it \textup{with}} diffusion $M_{\rm iso}$ as
\begin{equation}
\label{eq:MisowD}
  M_{\rm iso} = M_{\rm iso}^\dagger + \frac{\Pi_{\rm crit}}{\lambda} {\rm M}_{\rm E} \ .
\end{equation}
Using eq.~\ref{eq:MisowD}, we study in the next section the effect of this new-found pebble isolation mass on the formation of planetary systems and the core masses of the formed planets.

\section{Influence on planet formation}
\label{sec:formation}

The pebble isolation mass determines the final mass of the planetary core in the pebble accretion scenario because the reduced accretion luminosity facilitates the accretion of gas \citep{2014A&A...572A..35L, LL2017} and the planet can eventually grow to become a gas giant. The formation pathway of the growing planet is determined by the growth rate and size of the planetary core because it influences its gas accretion rates \citep{2014arXiv1412.5185P} and migration behaviour \citep{2013arXiv1312.4293B}. By reaching a different pebble isolation mass, the planet can undergo a different formation pathway.

We therefore investigate in this section the influence of the pebble isolation mass on planet growth by comparing planet growth simulations to the new pebble isolation mass (eq.~\ref{eq:MisowD}) with simulations with the pebble isolation mass measured by \citet{2014A&A...572A..35L}, who only inferred the dependence on $H/r$. For this we make use of the planet growth simulations including planet migration and disc evolution presented in \citet{2015A&A...582A.112B}.

\subsection{Planet growth and migration model}

The planet growth and migration model is described in great detail in \citet{2015A&A...582A.112B}, therefore we only repeat the essential points here. The planet growth and migration rates strongly depend on the disc structure. We used here the disc structure model of \citet{2015A&A...575A..28B}. This semi-analytical disc model features bumps and dips in the inner disc structure caused by transitions in the opacity at the water ice line, which can act as planet traps for low-mass planets \citep{2015A&A...575A..28B,2016A&A...590A.101B} and evolves in time. We used a disc lifetime of $3$ Myr.

The growth rate of the planet depends on the pebble surface density $\Sigma_{\rm peb}$ at the planet location of the protoplanetary disc
 \begin{equation}
 \label{eq:Mdotpebble}
  \dot{M}_{\rm c} = 2 \left(\frac{\tau_{\rm f}}{0.1}\right)^{2/3} r_{\rm H} v_{\rm H} \Sigma_{\rm peb} \ ,
 \end{equation}
where $r_{\rm H}$ is the planetary Hill radius and $v_{\rm H}$ the Hill speed at which the particles enter, given by $v_{\rm H} = r_{\rm H} \Omega_{\rm K}$. In the drift-limited growth of dust particles to pebbles \citep{2012A&A...539A.148B}, the pebble surface density depends on the pebble flux $\dot{M}_{\rm peb}$ \citep{2014A&A...572A.107L} in the following way:
 \begin{equation}
  \label{eq:SigmaPeb}
  \Sigma_{\rm peb} = \sqrt{\frac{2 \dot{M}_{\rm peb} \Sigma_{\rm g} }{\sqrt{3} \pi \epsilon_{\rm P} r_{\rm P} v_{\rm K}}} \ .
 \end{equation}
Here $\epsilon_{\rm P} =0.5$ \citep{2014A&A...572A.107L}. We
note that the nominal pebble flux used in \citet{2015A&A...582A.112B} was overestimated by a factor of $\approx$10 and that we used here a modified pebble growth model presented in Bitsch et al. (2017).

After the planet has reached its pebble-isolation mass, it can contract a gaseous envelope \citep{2014ApJ...786...21P}, and as soon as the mass of the gaseous envelope is higher than the core mass of the planet, runaway gas accretion can start \citep{2010MNRAS.405.1227M}.

Growing planets interact with their natal protoplanetary disc and migrate in it. Low-mass planets do not perturb the disc significantly and migrate in type-I migration, which depends mainly on the disc viscosity and on the radial gradients of surface density, temperature, and entropy \citep{2011MNRAS.410..293P}. The disc structure is therefore of crucial importance in determining the migration rates. Planets growing further (e.g. as a result of rapid gas accretion) push the gas away from their orbit (or accrete it, \citealt{2017Icar..285..145C}) and open a gap in the protoplanetary disc and migrate in type II migration. This migration rate depends on the viscosity of the protoplanetary disc. For a review on planet migration, see for example \citet{2013arXiv1312.4293B}.

To calculate the torque $\Gamma$ exerted by the disc on the planet, we followed \citet{2011MNRAS.410..293P} for type I migration and the viscous evolution for type II migration. The orbital migration time $t_{\rm m}$ is given as
\begin{equation}
 t_{\rm m} = - \frac{J}{(dJ/dt)} \ ,
\end{equation}
where $J$ is the angular momentum of the protoplanet. Migration here is thus defined in terms of the total torque exerted on the orbit. The migration time so defined is positive when the total torque is negative. For constant eccentricity, the time taken to migrate to the centre is $t_{\rm m}/2$ \citep{2000MNRAS.315..823P}. This factor of $2$ was absent in the original simulations of \citet{2015A&A...582A.112B} and was now added here.

\subsection{Stokes numbers in the planet formation model}

\begin{figure}
 \centering
 \includegraphics[scale=0.7]{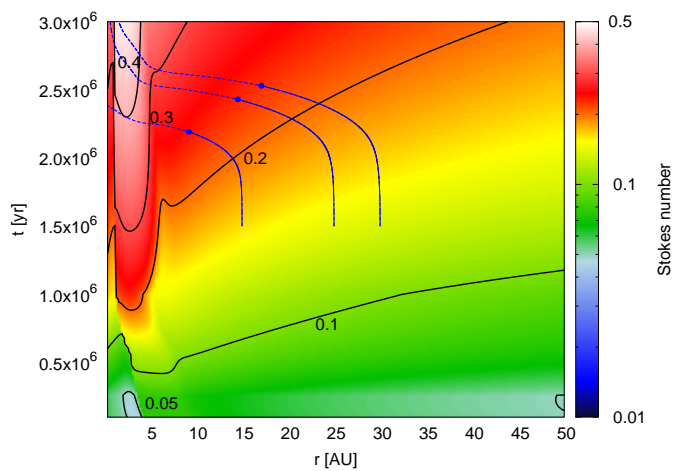} 
 \caption{Stokes number of the pebbles in the simulations presented in Figs.~\ref{fig:r0t0} and~\ref{fig:r0t0core} at a given time $t$ and orbital distance $r$ in the disc. The black lines mark Stokes numbers of $0.05$ to $0.4$. Note that each planet growth trajectory stars at a given point in $r$-$t$ and the planet the moves to higher time values, indicating that the Stokes number of the pebbles accreted by the planet increases. Additionally, the planet migrates in the disc to smaller orbital distances, which also increase the Stokes number of the accreted particles. The over-plotted blue lines correspond to the growth tracks shown in Fig.~\ref{fig:tracks}, where the solid line corresponds to solid (pebble) accretion, the dot marks the pebble-isolation mass, and the dashed line corresponds to the gas accretion phase. The planetary growth track moves upwards to increasing time.
   \label{fig:r0t0Stokes}
   }
\end{figure}

The Stokes numbers of the pebbles in our model are shown in Fig.~\ref{fig:r0t0Stokes} as a function of distance and time. The blue lines indicate the growth tracks of the planets shown in Fig.~\ref{fig:tracks} as the planets grow and migrate. The solid lines correspond to solid accretion, which stops when the planet reaches pebble-isolation mass (marked as a dot), while the dashed lines indicates gas accretion.
 
Typically, the pebbles accreted by planets in our model have Stokes numbers in the range of $0.05 < \tau_{\rm f} < 0.4$, as those are the Stokes numbers of the pebbles dominated by radial drift \citep{2012A&A...539A.148B, 2014A&A...572A.107L}. As shown in section~\ref{subsec:diffusion}, only small pebbles can drift through the pressure bump (with $\tau_{\rm f} < 0.01$ when the planet just reached pebble-isolation mass in the absence of diffusion, $M_{\rm iso}^\dagger$). In our model, $\alpha=0.0054$, indicating a strong turbulent diffusion efficiency (Fig.~\ref{fig:Stokesmap}), but even turbulent diffusion cannot carry the particles of the dominant Stokes number across the pressure bump because of the large pebble sizes (Fig.~\ref{fig:r0t0Stokes}). Therefore the pebble-isolation mass is only slightly increased by the effects of turbulent diffusion in our model. This means that the bulk of the pebbles and therefore also the bulk of the solid mass is blocked by a planet that has reached the pebble-isolation mass if the particle size is dominated by radial drift. Additionally, the blocking of pebbles becomes more efficient as the planet grows and starts to accrete a gaseous envelope.

The final pebble sizes in our model were determined by radial drift alone, where we did not take the effects of fragmentation \citep{2008A&A...480..859B, 2011A&A...525A..11B}, bouncing \citep{2010A&A...513A..57Z}, or condensation \citep{2013A&A...552A.137R, 2017A&A...602A..21S} into account. Fragmentation and bouncing can lead to smaller pebble sizes than in the drift-limited case, while condensation of volatiles onto already existing pebbles can increase their size. 

The sizes of pebbles determined by the fragmentation limit are given in \citet{2015ApJ...813L..14B} as
\begin{equation}
 a_{\rm f} = \frac{2}{3 \pi} \frac{\Sigma_{\rm g}}{\rho_{\rm peb} \alpha} \frac{v_{\rm f}}{c_{\rm s}} \ ,
\end{equation}
where $\rho_{\rm peb}$ is the density of the pebble itself (set to 1.5 g/cm$^3$ for water ice) and $v_{\rm f}$ is the fragmentation velocity limit, where water-ice particles have a higher fragmentation velocity of $10$ m/s than silicate grains \citep{2015ApJ...798...34G}. When we use only this fragmentation limit for icy particles, the Stokes numbers of the particles in our disc model are 0.01-0.2, which is slightly smaller than in the drift-limited case (shown in Fig.~\ref{fig:r0t0Stokes}). Lower Stokes numbers will result in higher pebble-isolation masses (eq.~\ref{eq:MisowD}) and thus higher core masses. However, condensation at ice lines could lead to even larger particles in these regions. Future models of planet formation have to take these effects into account in order to calculate more realistic grain sizes.

In the following, we use eq.~\ref{eq:MisowD} to include the effects of diffusion in the calculations of the pebble-isolation mass and to simulate the growth of planets through pebble accretion, where we use the Stokes numbers of the drift-limited solution shown in Fig.~\ref{fig:r0t0Stokes}.

\subsection{Growth tracks}

Growth and migration depend on the structure of the protoplanetary disc, where we follow the disc model of \citet{2015A&A...575A..28B}, and use a dust metallicity of $Z_{\rm dust}=0.5\%$ to set the disc opacity. The disc viscosity is $\alpha=0.0054$. The planetary growth rate depends crucially on the amount of available pebbles that can be accreted by the planet. In the remainder of the paper, we use $Z_{\rm peb} = 1.0\%$ as in \citet{2015A&A...582A.112B}. We set the disc lifetime to be $3$ Myr.

\begin{figure}
 \centering
 \includegraphics[scale=0.7]{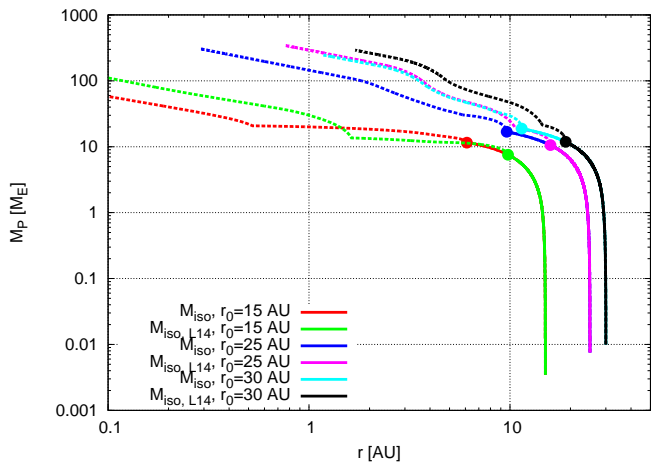} 
 \caption{Growth tracks of planets starting at several different initial positions in a disc that is already $1.5$ Myr old for pebble-isolation masses $M_{\rm iso}^{L14}$ given by \citet{2014A&A...572A..35L} in eq.~\ref{eq:Misolation} and for the new-found pebble isolation mass $M_{\rm iso}$ in eq.~\ref{eq:MisowD}. The initial growth phase is the same for both isolation masses, so the growth tracks diverge only when the pebble-isolation mass (eq.~\ref{eq:Misolation}) is reached, marked by the dots. The solid lines indicate pebble accretion, while the dashed lines mark gas accretion.
   \label{fig:tracks}
   }
\end{figure}

In Fig.~\ref{fig:tracks} we show the growth tracks of planetary seeds starting at different locations in a disc that is already $1.5$ Myr old, meaning that the planets evolve for $1.5$ Myr to reach a disc lifetime of $3$ Myr. For each orbital distance we ran two simulations, where the only difference was the final pebble-isolation mass, determined either by \citet{2014A&A...572A..35L}, eq.~\ref{eq:Misolation}, or by the new-found pebble-isolation mass, eq.~\ref{eq:MisowD}. This means that the initial growth is the same for the two simulations, until the planet in one simulation reaches the pebble-isolation mass and gas accretion starts. This is generally the case for simulations following the pebble-isolation mass of eq.~\ref{eq:Misolation}, which is generally lower than eq.~\ref{eq:MisowD}, especially for the given disc model with $\alpha=0.0054$.

The final core mass also determines the contraction phase of the envelope, where $\dot{M}_{\rm env,gas} \propto M_{\rm core}^{11/3}$ \citep{2014ApJ...786...21P}, which in turn determines how fast the planet transitions into runaway gas accretion and can then open a gap and transition into the slower type II migration phase.

In the inner parts of the protoplanetary disc, the pebble-isolation mass is low because of the low $H/r$, while the pebble-isolation mass is high in the outer parts of the disc because $H/r$ is high. For planets forming in the inner regions of the disc ($r<15$ AU), the difference between eq.~\ref{eq:Misolation} and eq.~\ref{eq:MisowD} is not that large, allowing planets to arrive at similar total masses and orbital distances. However, the core masses of planets formed using eq.~\ref{eq:MisowD} are higher. In the outer disc, the pebble flux is quite low, so that the differences in the pebble-isolation mass result in a slower growth of the planets, where $M_{\rm iso}$ is determined by eq.~\ref{eq:MisowD}, because pebble accretion is slower than gas contraction for these pebble densities. This also results in further inward migration before the planet opens a deep gap and transitions into type
II migration. In total, the differences regarding the final orbital position and the final planetary mass seem not very great. However, the differences in the core masses can be up to $30\%$ (see Fig.~\ref{fig:r0t0core}), which is crucial for the formation of the ice giants in our solar system, which had low core masses in \citet{2015A&A...582A.112B}.

\subsection{Global picture}

\begin{figure}
 \centering
 \includegraphics[scale=0.7]{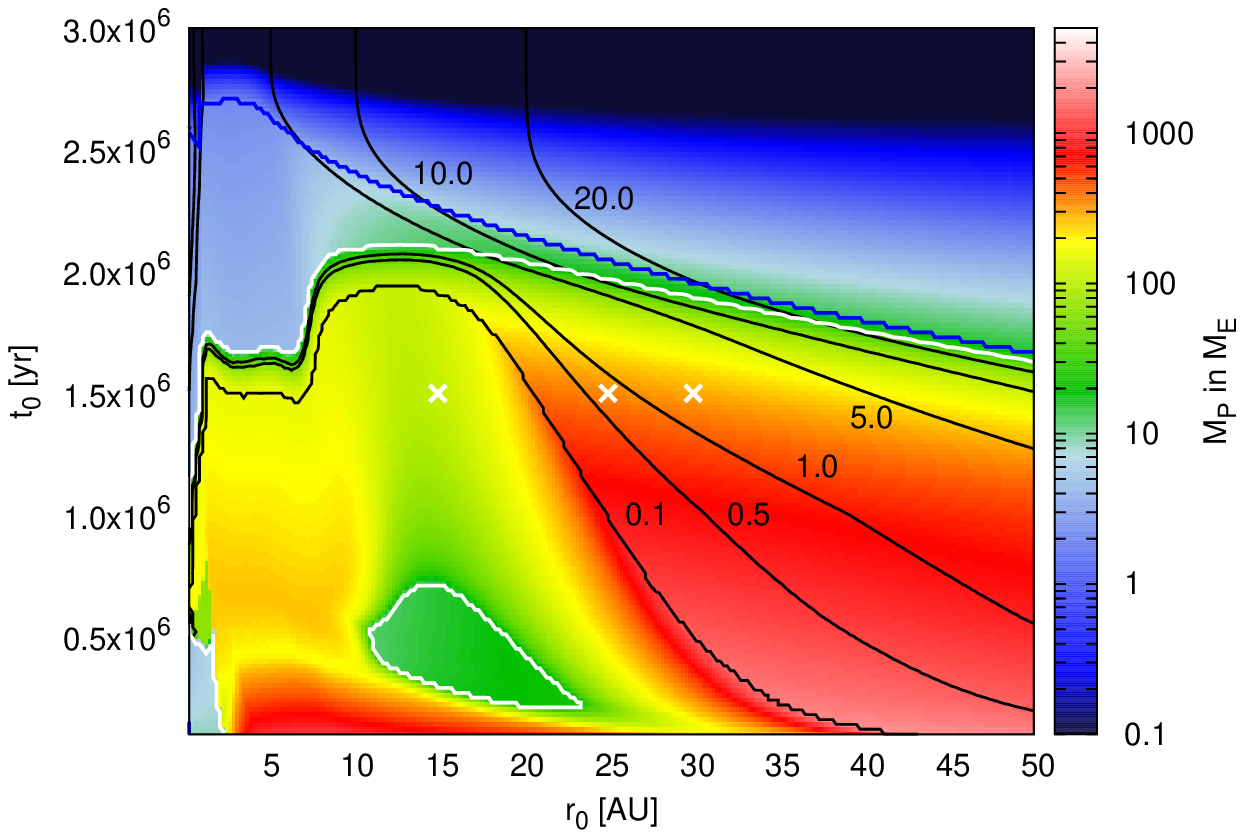} 
 \includegraphics[scale=0.7]{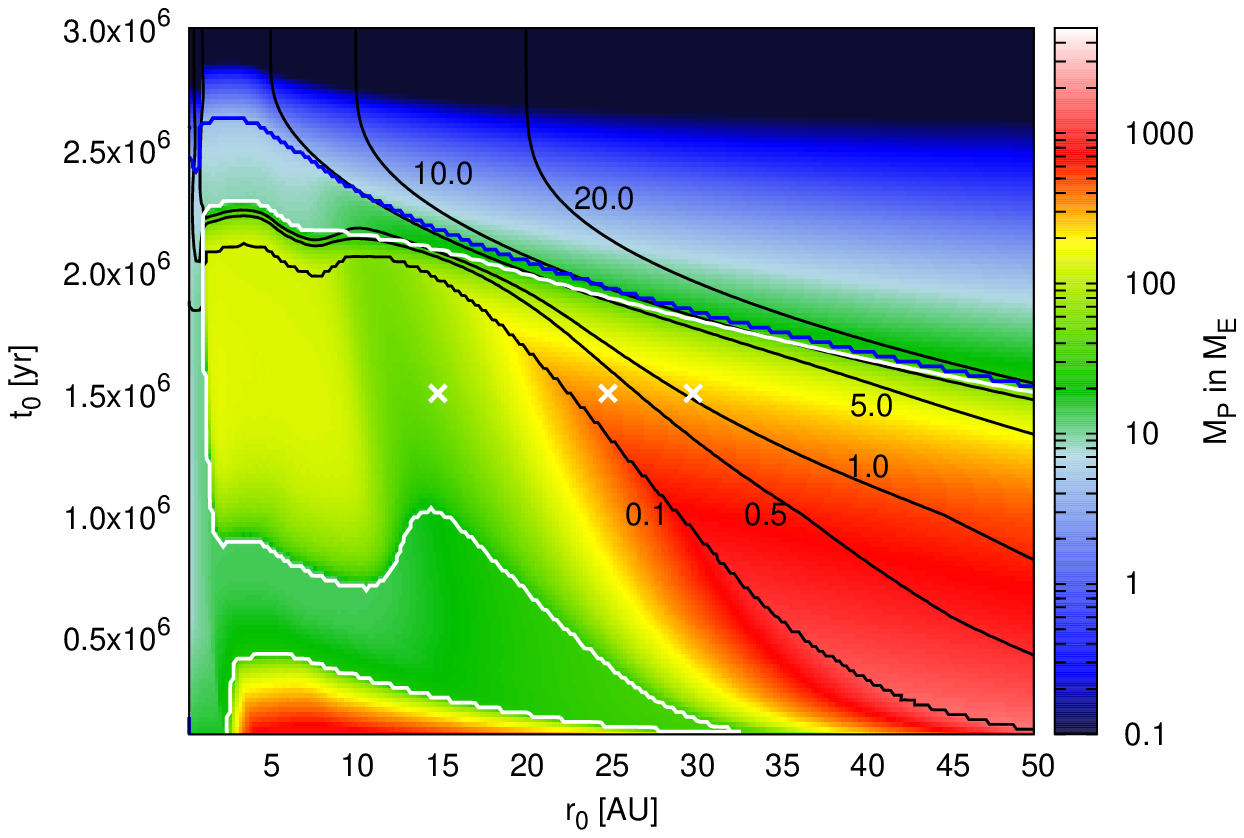}
 \caption{Final masses of planets (total mass $M_{\rm P}=M_{\rm c}+M_{\rm env}$) as a function of formation distance $r_{\rm 0}$ and formation time $t_{\rm 0}$ in the disc. Planets that are below the dark blue line have reached pebble-isolation mass and can accrete gas. All planets that are below the white line have $M_{\rm c}<M_{\rm env}$, indicating that they have undergone runaway gas accretion. The black lines indicate the final orbital distance $r_{\rm f}$ of the planet, namely $0.1$, $0.5$, $1.0$, $5.0$, $10.0,$ and $20.0$ AU. The top plot corresponds to the pebble-isolation mass found in \citet{2014A&A...572A..35L}, while the bottom plot corresponds to the pebble-isolation mass in this work (eq.~\ref{eq:MisoNEW}). The higher pebble-isolation mass suppresses the formation of gas giants in the very outer parts of the disc at late times.
   \label{fig:r0t0}
   }
\end{figure}

We now extend the approach of the growth tracks to probe the planetary growth for starting positions of the planetary seeds from $r_0=0.2$ AU to $50$ AU and from $t_0 = 100$ kyr to $3$ Myr. In Fig.~\ref{fig:r0t0} we show the final total planetary mass of planets as a function of $r_0$ and $t_0$ for pebble-isolation masses following eq.~\ref{eq:Misolation} (top) and eq.~\ref{eq:MisowD} (bottom). The white crosses mark the growth tracks shown in Fig.~\ref{fig:tracks}.

At first glance, the difference in final orbital position $r_{\rm f}$ and final planetary mass $M_{\rm p}$ is not that large compared for the different pebble- isolation masses, in agreement with Fig.~\ref{fig:tracks}. For $0.1$ Myr $<t_0<2.0$ Myr and $r<20$ AU, the formation of close-in planets that have reached the inner edge of the disc at $0.1$ AU is triggered. These planets form too close to the central star, so that planetary migration drives them towards the inner edge of the disc during the lifetime of the protoplanetary disc for our migration rates in discs with high viscosity. As the pebble-isolation mass is higher, the planetary cores with low-mass gaseous envelopes (that have not reached runaway gas accretion) become too large to be contained in the region of outward migration (which can only hold planets of a few Earth masses after about $1$ Myr; \citealt{2015A&A...575A..28B}). These planets then drift inwards as rock-dominated planets (bottom panel of Fig.~\ref{fig:r0t0} in contrast to the top panel of Fig.~\ref{fig:r0t0}). 

Planets forming in the outer part of the protoplanetary disc reach higher pebble-isolation masses owing to the higher aspect ratio. In the top panel of Fig.~\ref{fig:r0t0}, the pebble-isolation mass is reached earlier, but the core masses are high enough ($\approx$10 ${\rm M}_{\rm E}$) to allow a transition into runaway gas accretion and thus gas giant formation. However, in the bottom panel of Fig.~\ref{fig:r0t0}, the pebble-isolation mass is higher, which prolongs the core formation timescale and thus results in gas accretion at later stages. The overall differences in $r_{\rm f}$ and $M_{\rm p}$ are not very large, however.

\begin{figure}
 \centering
 \includegraphics[scale=0.7]{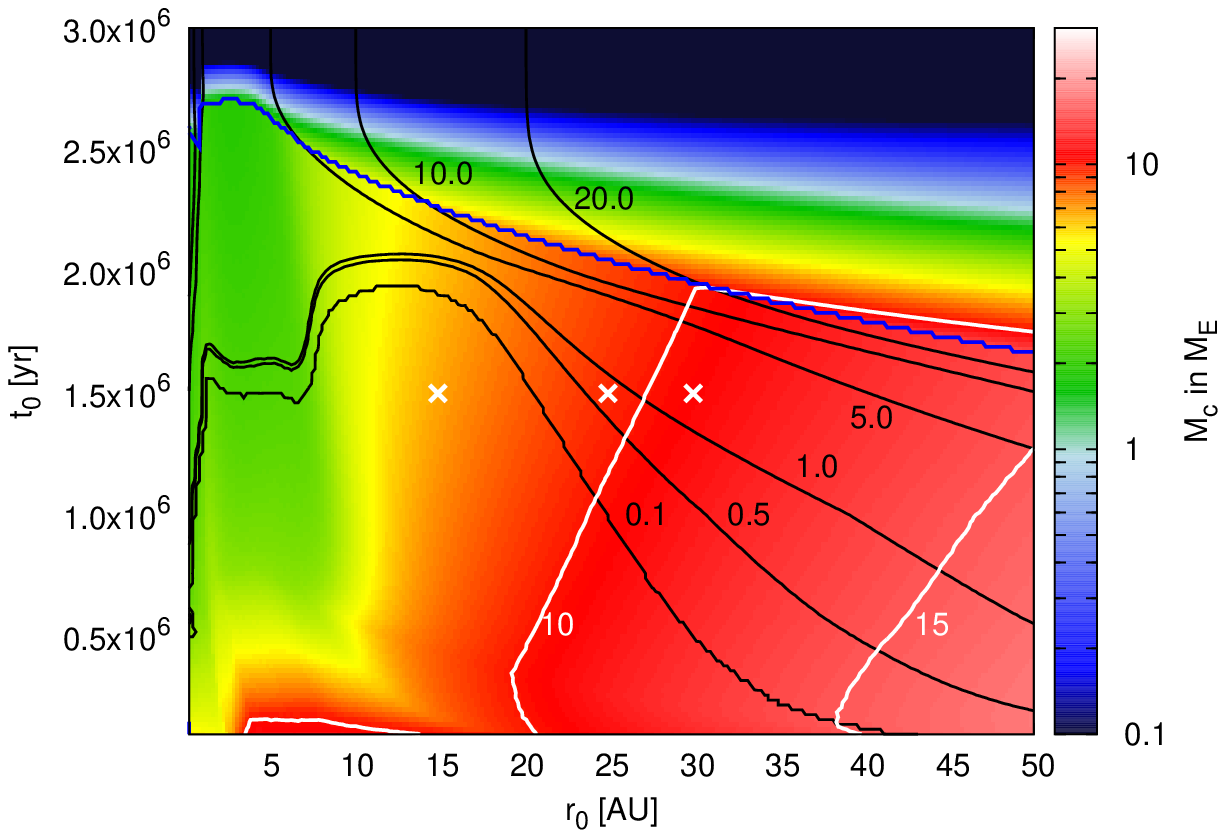} 
 \includegraphics[scale=0.7]{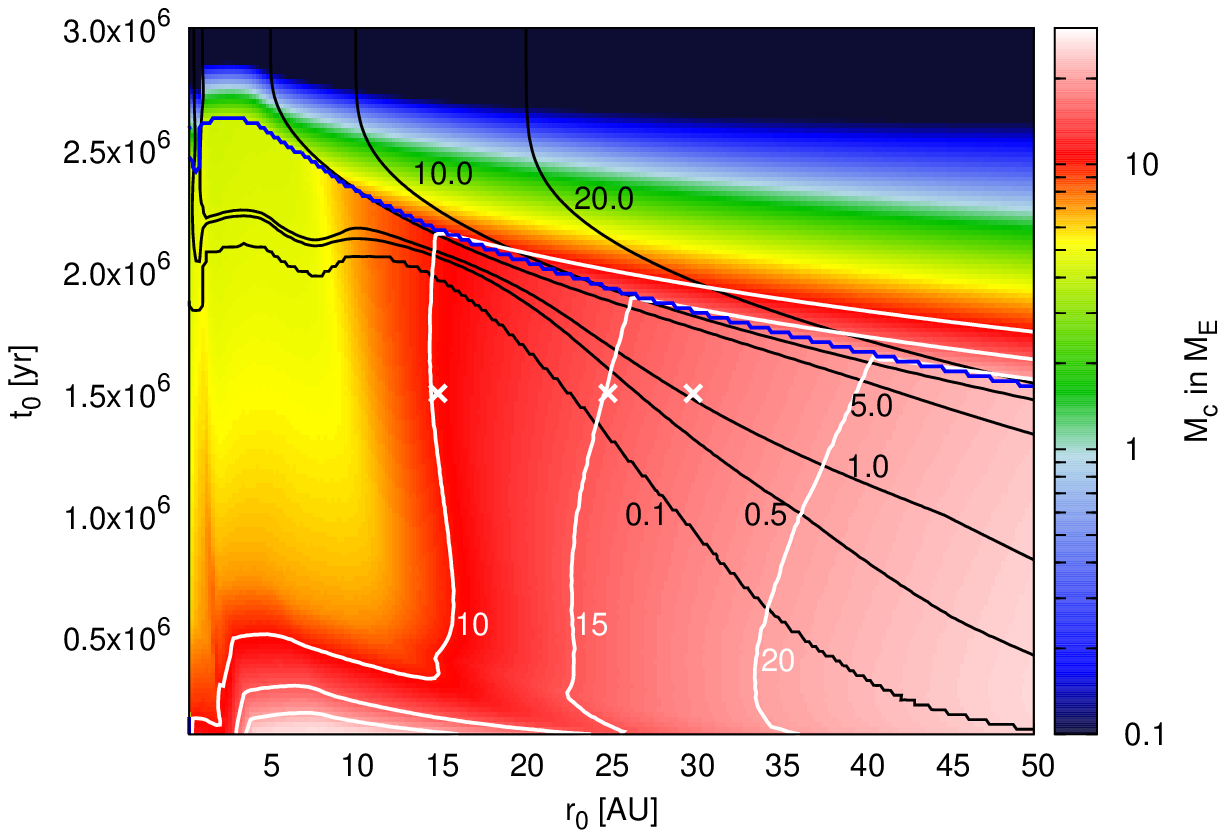}
 \caption{Final core masses of planets $M_{\rm c}$ as a function of formation distance $r_{\rm 0}$ and formation time $t_{\rm 0}$ in the disc. The white lines correspond to core masses of 10, 15, and 20 ${\rm M}_{\rm E}$ (top to bottom) and are marked by white numbers. The black and blue lines have the same meaning as in Fig.~\ref{fig:r0t0}. The top plot corresponds to the pebble-isolation mass stated in \citet{2014A&A...572A..35L}, while the bottom plot corresponds to the pebble-isolation mass in this work (eq.~\ref{eq:MisoNEW}). Clearly, eq.~\ref{eq:MisoNEW} allows for higher core masses, which is crucial for the formation of ice giants in our own solar system.
   \label{fig:r0t0core}
   }
\end{figure}

Clearer differences can be seen with respect to the core masses of these planets (Fig.~\ref{fig:r0t0core}), where eq.~\ref{eq:MisowD} delivers core masses that are about $\approx$30\% higher, compared to eq.~\ref{eq:Misolation} in our model with $\alpha=0.0054$. Figure~\ref{fig:Miso} shows the clear dependence on $H/r$ for the pebble-isolation mass, explaining why the core masses increase with orbital distance of the formed planets, because $H/r$ increases outwards in a stellar irradiated disc \citep{2013A&A...549A.124B, 2015A&A...575A..28B}. However, the $10$ ${\rm M}_{\rm E}$ core mass line seems to be constant as a function of $t_0$  at early times in the inner regions of the disc ($r<$ 15 AU) of Fig.~\ref{fig:r0t0core}: the growth time in the outer parts of the disc is much longer because of the lower pebble flux and the larger pebble scale height. The planetary growth additionally competes with planetary migration (driving the planet inwards to parts of the disc with lower $H/r$ and thus lower pebble-isolation mass) and disc evolution, where the disc aspect ratio decreases with time \citep{2015A&A...575A..28B}. 

These higher core masses are more consistent with the planetary structure of the solar system\footnote{The findings in \citet{2015A&A...582A.112B} and \citet{2016A&A...590A.101B} produced core masses around 10 ${\rm M}_{\rm E}$, approximately a factor 2 lower than Uranus and Neptune.}, where Uranus and Neptune have not reached the pebble-isolation mass \citep{2014A&A...572A..35L} and thus stayed at 15-20 ${\rm M}_{\rm E}$ without accreting a large gaseous envelope. The final core mass could additionally be increased if more pebbles are available, allowing a faster growth of the planets.

\section{Discussion}
\label{sec:discuss}

\subsection{Planet migration versus pebble-drift speeds}

Planets embedded in protoplanetary discs interact gravitationally with the disc and move through the disc \citep{2006A&A...459L..17P, 2008ApJ...672.1054B, 2009A&A...506..971K, 2011MNRAS.410..293P}. The migration timescale $\tau_{\rm mig}$ of low-mass planets is estimated in \citet{2002ApJ...565.1257T} and given as
\begin{eqnarray}
\label{eq:planetmig}
 \tau_{\rm mig} &=& C \frac{{\rm M}_\odot}{M_{\rm pl}} \frac{{\rm M}_\odot}{\Sigma_{\rm g}(r_{\rm pl}) r_{\rm pl}^2} \left( \frac{H}{r} \right)^2_{\rm pl} \Omega_K^{-1} \nonumber \\
  &\approx& 7.8 \times 10^5 \left( \frac{M_{\rm pl}}{{\rm M}_{\rm E}} \right)^{-1} {\rm yr} \ .
\end{eqnarray}
Here, $r_{\rm pl}$ and $(H/r)_{\rm pl}=0.05={\rm const.}$ are the orbital distance and the aspect ratio at the planet location. The constant $C$ reflects the migration speed through the disc surface density profile and disc temperature profile, given by $C=1 /(2.5+1.7 \beta_{\rm T} - 0.1 \alpha_\Sigma)$ \citep{2011MNRAS.410..293P}, where $\alpha_\Sigma$ is given by $\Sigma_{\rm g} = \Sigma_0 r^{-\alpha_\Sigma}$ with $\Sigma_0 =$ 350 g/cm$^2$ and $\beta_{\rm T}$ by $T \propto r^{-\beta_{\rm T}}$. For our standard disc with $\alpha_\Sigma = 0.5$ and $\beta_{\rm T} = 1,$ the pre-factor is $C=0.24$. 

We can now compare this with the radial drift speed of particles \citep{2008A&A...480..859B}, which is given by
\begin{equation}
 v_{\rm d,rad,tot} = v_{\rm r, d} + \frac{v_{\rm r,gas}}{1+\tau_{\rm f}^2} \ .
\end{equation}
The radial speed of the gas $v_{\rm r,gas}$ in an $\alpha$ disc is estimated by \citet{Takeuchi2002} as
\begin{equation}
 v_{\rm r,gas} = - 3 \alpha \frac{c_{\rm s}^2}{v_{\rm K}} \left( \frac{3}{2} - \alpha_\Sigma \right) \ .
\end{equation}
The quantity $v_{\rm r,gas}$ that describes the radial drift of individual dust particles is given by \citet{1977MNRAS.180...57W} as
\begin{equation}
 v_{\rm r, d} = - \frac{2 \Delta v}{\tau_{\rm f} + 1/\tau_{\rm f}} \ ,
\end{equation}
where $\Delta v$ is the maximum drift velocity, which is calculated as
\begin{equation}
 \Delta v = \frac{c_{\rm s}^2}{2 v_{\rm K}} \left(\alpha_\Sigma + \frac{7}{4} \right) \ .
\end{equation}

\begin{figure}
 \centering
 \includegraphics[scale=0.7]{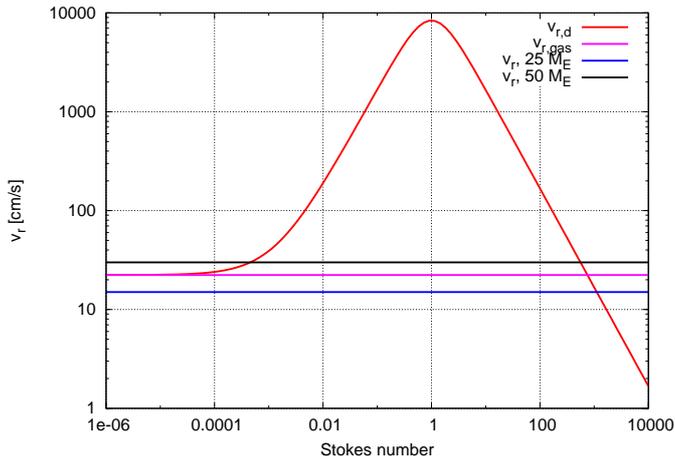} 
 \caption{Total inward velocity of a dust particle (red) as a function of Stokes number $\tau_{\rm f}$ and the corresponding gas velocity for a disc with $\alpha=0.001$ (magenta). Over-plotted are also the radial speeds of a 25 and a 50 ${\rm M}_{\rm E}$ planet in the same discs, assuming pure type I planetary migration. Even for the 50 ${\rm M}_{\rm E}$ planets, particles with $\tau_{\rm f} > 0.001$ drift faster than the planet and would thus accumulate at the pressure bump outside of the planetary orbit.
   \label{fig:ParticlePlanetdrift}
   }
\end{figure}

In Fig.~\ref{fig:ParticlePlanetdrift} we show the radial drift speed of particles as a function of Stokes number, the radial gas velocity, and the migration speed of planets with 25 ${\rm M}_{\rm E}$ (corresponding directly to the pebble-isolation mass without diffusion) and 50 ${\rm M}_{\rm E}$ in a disc with $\alpha=0.001$. Clearly, the particles drift faster than the planet migrates when the pebble-isolation mass is reached. Even for the 50 ${\rm M}_{\rm E}$ planet, particles with $\tau_{\rm f} > 0.001$ drift faster than the planet migrates, indicating that particles drifting inwards from the outer disc will be trapped in the pressure bump outside of the planetary orbit.

However, planetary migration is more complicated than the simple estimate provided in eq.~\ref{eq:planetmig}. The corotation torque can change the migration speed and also the direction of migration \citep{2006A&A...459L..17P, 2008ApJ...672.1054B, 2009A&A...506..971K, 2011MNRAS.410..293P}. This can result in regions of the disc where the planet does not migrate at all, so-called zero-torque regions \citep{2013A&A...555A.124B, 2014A&A...564A.135B, 2015A&A...575A..28B, 2015arXiv150303352B}. However, the corotation torque is prone to saturation, which depends on the local viscosity of the protoplanetary discs \citep{2011MNRAS.410..293P}, where a lower viscosity allows an easier torque saturation, preventing outward migration. Even in these cases, however, the pebbles with $\tau_{\rm f}> 0.001$ drift faster than the planet migrates and will thus accumulate outside of the planetary orbit in the generated pressure bump.

\subsection{Mass loading in the pressure bump}

As soon as the planet reaches the pebble-isolation mass, pebbles drifting inwards from the outer disc are stopped in the pressure bump generated by the planet. As the flux of pebbles from the outer disc continues, pebbles accumulate in the pressure bump and the pebble-to-gas ratio increases. However, an increased pebble-to-gas ratio will trigger the streaming instability \citep{2010ApJ...722L.220B, 2015A&A...579A..43C}, transforming the pebbles into planetesimals. For the streaming instability to occur, a vertically integrated pebble-to-gas ratio of a few percent is needed.

Pebbles can also accumulate in vortices generated outside of the gap carved by the planet, where they would form planetesimals. \citet{2015arXiv150105364R} showed that the accumulated pebbles could destroy a vortex in a disc, but in our case, the vortex is fed by the presence of the planet itself, which was not taken into account in their work. \citet{2018MNRAS.473..796A} studied the linear growth regime of the streaming instability in pressure bumps in discs, and they found that streaming instability can occur within the pressure bump. The accumulated pebbles inside the pressure bump therefore turn into planetesimals, which do not affect the gas velocities and thus do not disrupt the pressure bump. However, how the presence of a planet influences the streaming instability in a pressure bump is still subject to investigation.

Nevertheless, the current evidence occurrence of planetesimal formation inside the pressure bump before mass loading with pebbles influences the gas dynamics of the pressure bump itself. This makes the pressure bump outside of the planetary orbit an interesting candidate for subsequent planet formation.

\subsection{Particle filtering by proto-Jupiters}

The pebble accretion scenario does not only allow for fast accretion of planetary cores at large distances, it also gives potential solutions to (a) the great dichotomy between the terrestrial planets and the gas giants \citep{2015Icar..258..418M}, (b) the inward motion of the water ice line as the protoplanetary disc evolves in time and crosses the orbit of the Earth in less than 1 Myr \citep{2016Icar..267..368M}, and (c) explain the difference between the non-carbonaceous and carbonaceous meteorites through different isolated reservoirs \citep{2017LPI....48.1386K}. The solution to all these problems could be related to the growth of the Jupiter core and to the amount of pebbles that can drift past it after reaching pebble-isolation mass.

In these scenarios, the Jupiter core forms in the cold part of the protoplanetary disc ($r>r_{\rm ice}$), where the pebbles are large. This makes the accretion very efficient because larger pebbles can be accreted more efficiently, allowing Jupiter to grow faster than the bodies in the terrestrial region \citep{2015Icar..258..418M}. Additionally, as soon as Jupiter reaches its pebble-isolation mass, the inward flux of large pebbles ($\tau_{\rm f}> 10^{-1}$) is stopped. The small pebbles drifting through are accreted very inefficiently (unless they grow again through coagulation), slowing down the growth of the bodies in the terrestrial region.

\citet{2016Icar..267..368M} reported that the inward flux of icy pebbles was stopped by a proto-Jupiter that had reached pebble-isolation mass, thus fossilizing the water ice line at $\approx$3 AU because the bulk of the material was stopped outside of the proto-Jupiter. However, a small fraction of water ice is needed to explain ordinary chondrites (and to a lesser extent, even enstatite chondrites) as they show evidence for some water alteration. The amount of water available had to be much lower than expected from solar proportion, however. This shows that these meteorites formed in a cold environment, but the availability of icy grains was somehow strongly reduced. The passage of small grains ($<10\upmu$m) through the barrier at the Jupiter pressure bump coupled with the inefficient accretion of such small grains explains these observations and is in agreement with the blocking efficiency of planets at pebble-isolation mass found in this study.

\citet{2017LPI....48.1386K} showed through meteoritic evidence that the reservoir between non-carbonaceous meteorites and carbonaceous meteorites was spatially separated in the protoplanetary disc around the young Sun at about $\approx$1 Myr. This separation can be achieved by a growing planet that stops the inward flux of particles, which corresponds to the core of Jupiter. \citet{2017LPI....48.1386K} gave the mass of the Jupiter core as $\approx$20 ${\rm M}_{\rm E}$. This is in agreement with thepebble-isolation mass we found
here, but the exact mass at which a growing Jupiter generated a pressure bump outside of its orbit depends on the disc properties (viscosity, aspect ratio, and pressure gradient).

\subsection{Ice giant formation}

As soon as the planet reaches its pebble-isolation mass, the envelope of the planets is no longer heated by infalling pebbles, and gas accretion can start \citep{2014A&A...572A..35L}. This initial contraction of the gaseous envelope depends on the cooling for the envelope and with that on the opacity inside it \citep{2014arXiv1412.5185P}. At low temperatures (below $1000$ K), the opacity is dominated by the dust grains, where a larger amount of dust grains increases the opacity and thus prolongs the contraction of the planetary envelope. However, in the pebble accretion scenario, the planet blocks the influx of new pebbles, and only very tiny dust grains can reach the planet. 

As soon as the main flux of pebbles onto the planet is stopped, the core stops to grow, but very small grains might still enter the planetary atmosphere and thus keep the opacity high, prolonging gas envelope contraction \citep{LL2017}. However, even in the case of no diffusion, particles with $\tau_{\rm f}<0.005$ can reach the planet (Fig.~\ref{fig:DriftM25}) and keep the opacity high. Only for larger planets can the pebble flux sufficiently be reduced. Additionally, this depends on the viscosity of the protoplanetary disc because of the diffusion of particles through the pressure bump, where higher viscosities allow a more efficient diffusion and the planet has to reach higher masses to block pebbles with the same Stokes number compared to planets in low-viscosity environments. This initial growth stage might be very important for the growth of ice giants, preventing them from immediately entering into rapid gas accretion and thus explain why Uranus and Neptune have envelopes of $10-15\%$ of their mass without entering into runaway gas accretion.

\citet{2014A&A...572A..35L} found that Uranus and Neptune may never have reached pebble-isolation mass and that this prevented them for accreting gas. Our findings indicate that these planets might have reached the pebble-isolation mass without diffusion limit (to stop efficient growth of the core), but the influx of small particles prevented an efficient cooling of the atmosphere and thus runaway gas accretion. However, this could only have
occurred if the viscosity of the protoplanetary disc was very low ($\alpha \approx$10$^{-4}$) because otherwise the pebble-isolation mass is higher than the mass of Uranus and Neptune for typically expected disc aspect ratios ($H/r>0.04$) in the outer disc at late disc evolution stages. We note that the outer disc structure is dominated by stellar irradiation, so that it is independent of the disc viscosity and $H/r$ becomes smaller through the reduced stellar irradiation as the system ages \citep{2015A&A...575A..28B}.

\section{Summary}
\label{sec:summary}

We have conducted 3D hydrodynamical simulations to measure the pebble-isolation mass as a function of the disc structure and turbulence strength. In particular, we investigated the dependence on the disc aspect ratio $H/r$, the disc viscosity (parametrised through $\alpha$), the radial pressure gradient $\partial \ln P / \partial \ln r, $  and the particle size described by the Stokes number $\tau_{\rm f}$. Our findings generally agree with the results presented in \citet{2014A&A...572A..35L}, who inferred the cubic dependence on the disc aspect ratio $H/r$, but we refined the pebble-isolation mass to more disc parameters and also confirmed our results in fully radiative discs with heating and cooling. In eq.~\ref{eq:MisowD} we provide the pebble-isolation mass as a function of our investigated disc parameters, which is useful for planet formation simulations involving pebble accretion \citep{2015A&A...582A.112B, 2015Natur.524..322L, 2016ApJ...825...63C, 2017A&A...607A..67M}.

Our findings result in a pebble-isolation mass that is up to a factor of $2-3$ higher than found in \citet{2014A&A...572A..35L}in the high-viscosity case ($\alpha \sim 0.01$)  and a factor of $\approx$1.5 higher than in 2D simulations, see Appendix~\ref{ap:2Ddiscs}. A higher viscosity and a steeper radial pressure gradient both result in a higher pebble-isolation mass. For very low viscosities, our simulations match the results of \citet{2014A&A...572A..35L}. This results in higher core masses in planet formation simulations compared to previous simulations (section~\ref{sec:formation}) in discs with $\alpha=0.0054$. Discs with higher viscosity thus better match the heavy element content of the ice giants in our own solar system, because even as the disc evolves and $H/r$ decreases, the pebble isolation mass stays high enough so that the ice giants did not reach pebble isolation mass. Additionally, discs with higher viscosity can more easily match overall the heavy element content of giant exoplanets \citep{2016ApJ...831...64T}.

We also investigated the penetration of particles of various Stokes number $\tau_{\rm f}$ through the pressure bump by radial advection with the gas and through turbulent diffusion. In the {\it \textup{absence of turbulent diffusion}}, a planet that has reached pebble-isolation mass can readily block pebbles with $\tau_{\rm f} > 0.005$, while a mass higher by a factor two is necessary to block pebbles with Stokes numbers as low as $\tau_{\rm f} \sim 10^{-4}$. {\it \textup{Including turbulent diffusion}} of particles due to viscosity changes this picture. Depending on viscosity and the particle size, the generated pressure bump needs to be stronger. To block particles of $\tau_{\rm f}=0.05$, a typical size in drift-limited pebble growth models \citep{2012A&A...539A.148B}, the planetary mass has to be increased by up to a factor of $\approx$2 compared to the pebble-isolation mass without turbulent diffusion (eq.~\ref{eq:MisoNEW}) for high-viscosity discs. In low-viscosity discs, blocking of particles with $\tau_{\rm f}=0.05$ requires a much smaller increase of the planetary mass than in the pebble-isolation mass without diffusion (Fig.~\ref{fig:Stokesmap}).

\begin{acknowledgements}

B.B. was supported by the Knut and Alice Wallenberg Foundation (grant 2012.0150). A.J.\, thanks the Knut and Alice Wallenberg Foundation (grants 2012.0150, 2014.0017, 2014.0048), the Swedish Research Council (grant 2014-5775) and the European Research Council (ERC Consolidator Grant 724687-PLANETESYS) for their financial support. The computations were done on the “Mesocentre SIGAMM” machine, hosted by the Observatoire de la C\^{o}te d'Azur. We thank P. Pinilla for discussions of the diffusion of dust particles. We thank the referee John Chambers for the comments that helped to improve this work.

\end{acknowledgements}

\appendix
\section{Azimuthal disc structure}
\label{ap:structure}

\begin{figure}
 \centering
 \includegraphics[scale=0.69]{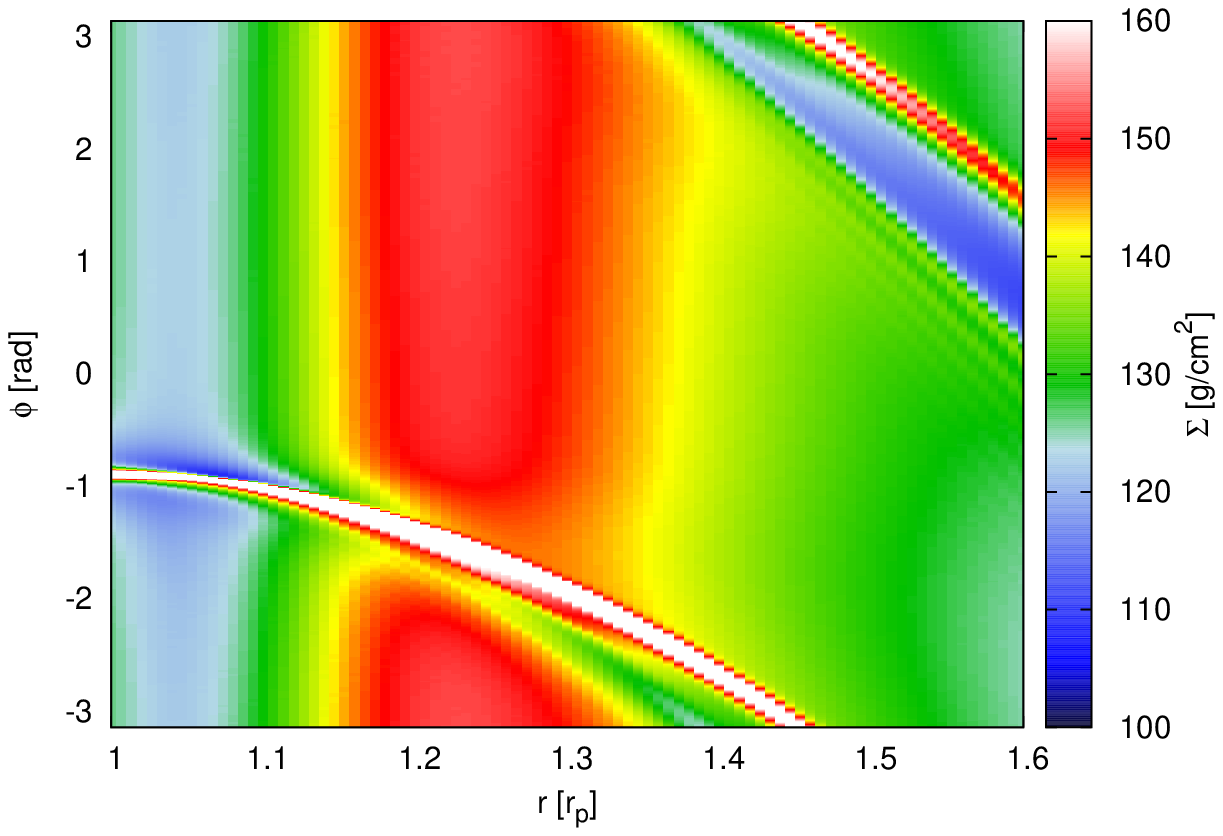} 
 \includegraphics[scale=0.69]{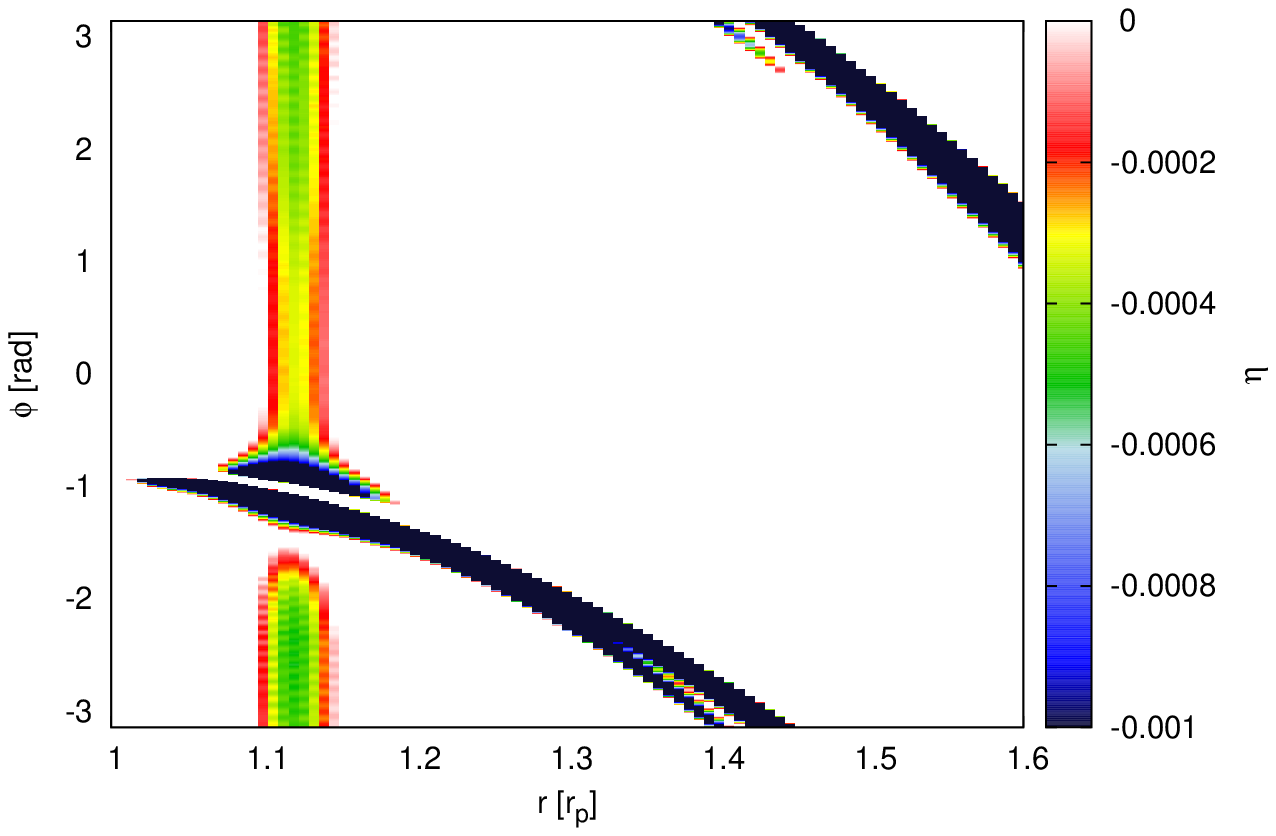} 
 \caption{Surface density (top) and $\eta$ value (bottom) for discs with $\alpha=0.001$, $H/r=0.05$ with an embedded 25 ${\rm M}_{\rm E}$ planet. The planet is located at $r=1$ and $\phi=-1$. A planet with this mass opens a partial gap in the disc density and has reached the pebble-isolation mass (fig.~\ref{fig:etaA001}). The $\eta$ value reaches negative values outside of the planetary orbit over the whole azimuthal range, indicating that taking the azimuthally average $\eta$ value to compute the pebble-isolation mass is correct.
   \label{fig:2Dmaps}
   }
\end{figure}

We present here the 2D surface density structure of a 25 ${\rm M}_{\rm E}$ planet embedded in a disc with $\alpha=0.001$ and $H/r=0.05$ (top panel of fig.~\ref{fig:2Dmaps}) as well as the $\eta$ value in the 2D configuration. The data of these simulations correspond to the purple line in Fig.~\ref{fig:etaA001}, indicating that the pebble-isolation mass has been reached when calculating the azimuthally averaged $\eta$ profile. However, as can be seen in the bottom panel of Fig.~\ref{fig:2Dmaps}, a negative $\eta$ value and with it a blocking of inward-drifting pebbles can be achieved at all azimuthal values when the pebble-isolation mass is reached. In principle, there is a small region in parameter space that might allow pebbles to ``tunnel'' through the pressure barrier, but the planet and with it the spiral waves orbit the central star at a frequency of $\Omega_P$, whereas the pebbles orbit with a frequency $\Omega(r)$, which results in the trapping of the pebbles inside the pressure bump.

\section{Comparison to 2D simulations}
\label{ap:2Ddiscs}

The pressure in isothermal 3D simulations is related to the gas volume density $P_{\rm 3D} = c_{\rm s}^2 \rho_{\rm g}$, while for isothermal 2D simulations, the pressure is related to the gas surface density with $P_{\rm 2D} = c_{\rm s}^2 \Sigma_{\rm g}$. However, the gas surface density and the gas volume density are related through
\begin{equation}
 \rho_{\rm g} = \frac{\Sigma_{\rm g}}{\sqrt{2 \pi} H_{\rm g}} \ ,
\end{equation}
where $H_{\rm g}$ is the gas scale height. For power-law discs in 3D with $\Sigma_{\rm g} \propto r^s$ and with $H/r={\rm const.}$, this implies $\rho_{\rm g} \propto r^{s-1}$ and $c_{\rm s}^2 \propto r^{-1}$. In 2D discs, the surface density needs to have $s>1$ to invert the pressure, while in 3D $s>2$ is needed, making it harder to open a pressure bump in the protoplanetary disc. 

\begin{figure}
 \centering
 \includegraphics[scale=0.7]{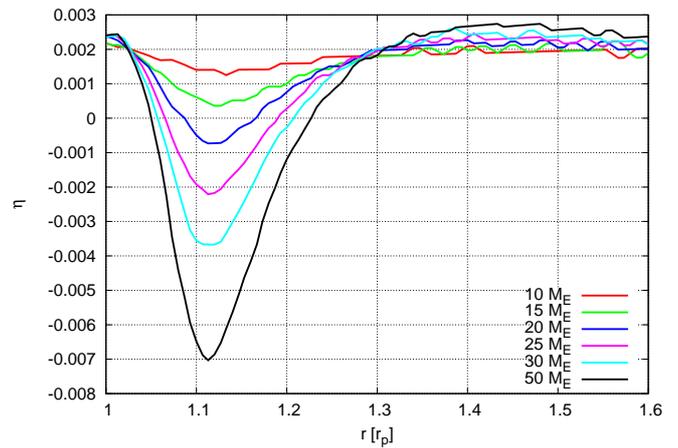}
 \caption{$\eta$ parameter as function of orbital distance from the planet for $\alpha=0.001$, $H/r=0.05$ and different planetary masses in 2D discs. The location of the planet is always at $r=1$. A negative $\eta$ parameter indicates the pressure bump. Here a mass of about $\approx$10 Earth masses is needed to generate the pressure bump, which is about a factor of $1.5-2$ lower than in the 3D case.
   \label{fig:eta2D}
   }
\end{figure}

In Fig.~\ref{fig:eta2D} we show the $\eta$ value as function of orbital distance for 2D discs with $\alpha=0.001$, $H/r=0.05,$ and different planetary masses, the same as in Fig.~\ref{fig:etaA001}. Clearly, a much lower planetary mass is needed to generate a pressure bump outside of the planetary orbit. In 2D simulations, planetary masses of about a factor of $1.5$ less are needed to open a pressure bump in the protoplanetary disc than in 3D simulations.

\bibliographystyle{aa}
\bibliography{Stellar}
\end{document}